\documentclass[%
aip,
amsmath,amssymb,
 preprint,
]{revtex4-1}

\usepackage{graphicx}
\usepackage{dcolumn}
\usepackage{bm}
\usepackage[utf8]{inputenc} 
\usepackage[T1]{fontenc}
\usepackage{CJKutf8}

\usepackage{mathrsfs}

\usepackage{etoolbox}

\usepackage[colorlinks=true, citecolor=blue]{hyperref}

\begin{document}

\begin{CJK*}{UTF8}{gbsn}

\title{Fourier neural operator for  large eddy simulation of compressible Rayleigh-Taylor turbulence}
\author{Tengfei Luo (罗腾飞)}
\author{Zhijie Li (李志杰)}
\affiliation{Department of Mechanics and Aerospace Engineering, Southern University of Science and Technology, Shenzhen 518055, China}%
\affiliation{Guangdong-Hong Kong-Macao Joint Laboratory for Data-Driven Fluid Mechanics and Engineering Applications, Southern University of Science and Technology, Shenzhen 518055, China}%
\affiliation{Guangdong Provincial Key Laboratory of Turbulence Research and Applications, Department of Mechanics and Aerospace Engineering, Southern University of Science and Technology, Shenzhen 518055, China}

\author{Zelong Yuan (袁泽龙)}
\affiliation{Harbin Engineering University Qingdao Innovation and Development Base, Qingdao 266000, China}%

\author{Wenhui Peng (彭文辉)}
\affiliation{The Hong Kong Polytechnic University, Department of Applied Mathematics, 999077, Hong Kong Special Administrative Region of China}%

\author{Tianyuan Liu (刘天源)}
\affiliation{College of Engineering, Peking University, Beijing, 100091, China}%

\author{Liangzhu (Leon) Wang}
\affiliation{Concordia University, Centre for Zero Energy Building Studies, Department of Building, Civil and Environmental Engineering, Montreal, H3G 1M8, Canada}%

\author{Jianchun Wang (王建春)}
\email{wangjc@sustech.edu.cn}
\affiliation{Department of Mechanics and Aerospace Engineering, Southern University of Science and Technology, Shenzhen 518055, China}%
\affiliation{Guangdong-Hong Kong-Macao Joint Laboratory for Data-Driven Fluid Mechanics and Engineering Applications, Southern University of Science and Technology, Shenzhen 518055, China}%
\affiliation{Guangdong Provincial Key Laboratory of Turbulence Research and Applications, Department of Mechanics and Aerospace Engineering, Southern University of Science and Technology, Shenzhen 518055, China}

\date{\today}

\begin{abstract}
The Fourier neural operator (FNO) framework is applied to the large eddy simulation (LES) of three-dimensional compressible Rayleigh-Taylor (RT) turbulence with miscible fluids at Atwood number $A_t=0.5$, stratification parameter $Sr=1.0$, and Reynolds numbers $Re=10000$ and 30000. The FNO model is first used for predicting three-dimensional compressible turbulence. The different magnitudes of physical fields are normalized using root mean square values for an easier training of FNO models. In the \emph{a posteriori} tests, the FNO model outperforms the velocity gradient model (VGM), the dynamic Smagorinsky model (DSM), and implicit large eddy simulation (ILES) in predicting various statistical quantities and instantaneous structures, and is particularly superior to traditional LES methods in predicting temperature fields and velocity divergence. Moreover, the computational efficiency of the FNO model is much higher than that of traditional LES methods. FNO models trained with short-time, low Reynolds number data exhibit a good generalization performance on longer-time predictions and higher Reynolds numbers in the \emph{a posteriori} tests.
\end{abstract}

\maketitle

\end{CJK*}

\section{\label{sec:int}INTRODUCTION}
Rayleigh-Taylor (RT) turbulence plays an important role in a variety of natural phenomena and industrial applications\cite{Zhou2017a}, including supernova explosion \cite{Hillebrandt2000,Burrows2000,Gamezo2003} and inertial confinement fusion (ICF) \cite{ATZENI2004,Petrasso1994,Betti2016}. Direct numerical simulation (DNS) is unrealistic for predicting RT turbulence at very high Reynolds numbers and complex situations. Large eddy simulation (LES) and Reynolds-averaged Navier-Stokes (RANS) are two major methods to reduce the computational cost of turbulence simulations. 

The RANS can provide mean-field information by modeling the Reynolds stresses, which have been widely studied and used in the RT, Richtmyer-Meshkov (RM), and Kelvin-Helmholtz (KH) turbulence.
\cite{Zhou2017b,Zhang2020jfm,Schilling2021,Morgan2022,Xiao2021,Kokkinakis2020} 
The LES simulates the large-scale motions by modeling the sub-grid scale (SGS) terms, which is more accurate than the RANS and computationally cheaper than DNS. Implicit large eddy simulations (ILES) for solving Euler or Navier-Stokes (NS) equations have been widely used in turbulence mixing,\cite{Dimonte2004pof} including TURMOIL3D code written by Youngs \cite{Youngs1991,Youngs1994}, RTI-3D code developed by Andrews,\cite{Andrews1995} and Miranda code developed at Lawrence Livermore National Laboratory.\cite{Cook2007,Cook2009,Cook2004}
Some explicit LES methods based on SGS models have also been used for the study of RT turbulence.\cite{Mellado2005,Nicoud1999,Burton2011,Mattner2004,ZhouH2021,Luo2023} 

Previous LES studies have focused more on statistics related to mixing and velocity fields, with less attention paid to thermodynamic quantities. However, compressibility and internal energy play an important role in the mixing and energy transfer in compressible RT turbulence.\cite{Luo2021,Luo2022,Zhao2020} The prediction of thermodynamic statistics cannot be ignored in LESs.\cite{Luo2023} \citet{Luo2023} gave direct comparisons of DNS, ILES, constant-coefficient spatial gradient model (CSGM), dynamic Smagorinsky model (DSM), and dynamic mixed model (DMM). They showed that structural models are significantly better than functional models and ILES, due to the fact that they can reasonably reconstruct small bubble and spike structures. The prediction of the temperature field is difficult for LES. The prediction results of the CSGM model are much better than those of the ILES, DSM, and DMM models, but still have a significant deviation from that of the filtered DNS (fDNS). The main reason is that the differences in the large-scale pressure-dilatation term calculated using DNS grids ($1024^3$) and using LES grids ($128^3$) are large, while the fluctuation of the temperature field is relatively small and is easier to be affected by the numerical errors of the pressure-dilatation term.\cite{Luo2021,Luo2022,Cook2001} This problem might be difficult to solve through traditional SGS models, and the newly developed neural operator methods would be a better choice.

With the development of deep learning (ML) technology, neural networks (NNs) have been widely applied to improve or replace traditional computational fluid dynamics (CFD) methods in turbulence simulations.\cite{brunton2020machine,duraisamy2019turbulence}
Various methods focus on using NNs to learn closure models for Reynolds stresses in RANS and SGS stresses in LES.\cite{sarghini2003neural,yuan2021dynamic,gamahara2017searching,guan2022stable,yang2019predictive,park2021toward,ling2016reynolds,tabe2023priori,wang2018investigations,Xie2019pre,Xie2020,xie2020artificial,Xu2023,beck2019deep,Maulik2019,Huang2019,Huang2021,Qi2022,Yin2022,Meng2023,li2021data}
Deep neural networks (DNNs) exhibit excellent performance in approximating highly nonlinear functions.\cite{lecun2015deep} Some pure data-driven methods aim to approximate the complete NS equations using DNNs.\cite{lusch2018deep,sirignano2018dgm,goswami2022deep,mohan2020spatio,li2024scalable}
Once trained, ``black-box'' NN models can be rapidly deployed on modern computers, achieving much higher efficiency compared to traditional CFD methods. Some studies explored incorporating physical information into DNNs. \cite{cai2021physics,lanthaler2022error,karniadakis2021physics,raissi2019physics,wang2020towards,jin2021nsfnets,xu2021explore}

Most DNN architectures aim to learn nonlinear mapping between finite-dimensional Euclidean spaces\cite{raissi2019physics,wu2020data,xu2021deep}. 
They excel at learning flow fields for specific flow states, but lack sufficient generalization ability for flow parameters, initial conditions, or boundary conditions.
\citet{li2020fourier} proposed the Fourier neural operator (FNO) method, which can effectively learn the mapping between infinite-dimensional spaces of input-output pairs. The FNO method learns the entire family of PDEs. The FNO method mimics the pseudospectral method, parametrizing the integral kernels in Fourier space, thus directly learning the mapping from function parameters to solutions. Benefiting from its rich expressive power and efficient architecture, the FNO model outperforms previous U-Net\cite{chen2019u}, TF-Net\cite{wang2020towards}, and ResNet\cite{he2016deep}. In low Reynolds number two-dimensional (2D) turbulence, the FNO method achieves a prediction error rate of 1\%.\cite{li2020fourier}  \citet{Meng2023} applied the FNO method to rapid prediction of the dynamic stall of 2D airfoils, also yielding excellent results. \citet{Li2022} applied the FNO model to LES of three-dimensional (3D) homogeneous isotropic turbulence (HIT), and its predictive performance was significantly better than traditional DSM and DMM models.

After the FNO method was proposed, it was extensively studied, and subsequently, some improved models were developed.\cite{Wang2024,li2023fourier,deng2023temporal,lehmann2023fourier,lyu2023multi,Peng2024Fourier,qin2024better} In 2D turbulence, \citet{Peng2022} introduced an attention-enhanced FNO model that can accurately reconstruct the instantaneous structures and statistics of 2D turbulence at high Reynolds numbers, further enhancing the prediction accuracy of the FNO model. Subsequently, they successfully applied it to 3D turbulence by linear attention approximation. \cite{Peng2023} The linear attention coupled FNO method (LAFNO) demonstrates excellent predictive performance in 3D HIT and free shear turbulence. \citet{wen2022u} developed a U-Net enhanced FNO method (U-FNO), which exhibits superior accuracy and efficiency in multi-phase flow problems. \citet{you2022learning} proposed an implicit FNO method (IFNO), modeling the increment between layers as integral operators to capture long-term dependencies in feature spaces. Combining the advantages of U-FNO and IFNO, \citet{Li2023} proposed an implicit U-Net enhanced FNO method (IU-FNO) for LES of turbulence. IU-FNO achieves efficient and accurate long-term predictions of turbulence, demonstrating a good performance in 3D HIT, free shear turbulence\cite{Li2023}, and turbulent channel flows\cite{Wang2024}. 
\citet{guibas2021adaptive} developed an adaptive FNO method (AFNO). AFNO splits high-resolution flow fields into several small local flow fields and uses FNO for block-wise prediction, addressing issues such as low training efficiency and insufficient GPU memory caused by high degrees of freedom. AFNO was employed by NVIDIA in global weather forecasting systems, achieving nearly a five-fold improvement in prediction efficiency compared to traditional methods while maintaining the same prediction accuracy. \citet{Li2024} introduced a transformer-based neural operator (TNO) to achieve precise and efficient predictions for LES of 3D HIT and free shear turbulence. The TNO model can achieve more stable long-term predictions and has fewer parameters than the FNO model.

Although NNs have been extensively studied in the simulation of turbulence, there is still little relevant research on RT turbulence. \citet{Gao2021} utilized a non-intrusive reduced order model (ROM) combining a proper orthogonal decomposition (POD) and an artificial neural network (ANN) to predict single-mode RT instability. \citet{Qiu2022} applied a physics-informed neural networks for phase-field method (PF-PINNs) to the problem of 2D single-mode RT instability, accurately predicting the growth of bubble and spike heights. \citet{Singh2019} embedded data-augmented models into RANS in RT turbulence, showing superior predictions of mixing heights and molecular mixing compared to traditional $k-L$ models. \citet{Xiao2023} applied physics-informed neural networks (PINNs) to solve RANS equations for one-dimensional (1D) RT turbulent mixing problems, and the PINNs method showed better prediction results than traditional $k-L$ models. \citet{XieH2023} developed a data-driven nonlinear $k-L$ model, applying the gene expression programming (GEP) method to provide an explicit and interpretable model that is easily ported to different RANS solvers. They trained the model using data from 2D tilted RT turbulent mixing, showing good predictive performance in 1D RT mixing, 1D reshocked RM mixing, quasi-1D KH mixing, and spherical implosion mixing. The model has good generalization performance and robustness, although it is solely trained by a single case.
Overall, NN methods for predicting 3D RT turbulent mixing is largely unexplored.

The FNO model has been shown to perform well in predicting turbulence problems, especially for complex 3D turbulence. \cite{Li2022,Li2023,Peng2023,Wang2024} As a purely data-driven model, it avoids some inherent shortcomings of SGS models and may perform well in compressible RT turbulence with miscible fluids. This study is the first attempt to apply the FNO method for predicting 3D compressible turbulence.
The rest of this paper is organized as follows. The next section presents the governing equations of DNS and LES, the SGS models, and computational methods. Section \ref{sec:FNO} provides the architecture of Fourier neural operator. Section \ref{sec:data} describes the dataset used for training the FNO model and the settings of the FNO model. Section \ref{sec:result} gives the \emph{a posteriori} tests. Finally, the main conclusions are summarized in \ref{sec:conclusion}.

\section{\label{sec:gq}GOVERNING EQUATIONS and NUMERICAL SIMULATIONS} 

A set of reference scales can be introduced to normalize the hydrodynamic and thermodynamic
variables: the reference length $L_r=L_x$, time $t_r=(L_x/g)^{1/2}$, velocity $u_r=(L_xg)^{1/2}$, density $\rho_r=(\rho_{H,0}+\rho_{L,0})/2$, temperature $T_r$, pressure $p_{r}=R\rho_rT_r/M_r$, concentration $c_r$, dynamic viscosity coefficient $\mu_r$, thermal conductivity coefficient $\kappa_{r}$, diffusion coefficient of species $D_r$, specific heat at a constant volume $C_{v,r}$ and molar weight $M_r=(M_H+M_L)/2$. Subscripts $H$ and $L$ represent the heavy and light fluids, respectively. $L_x$ is the horizontal length of the flow domain. $g$ is the acceleration imposed opposite to the vertical direction.
$\rho_{H,0}$ and $\rho_{L,0}$ represent the densities on both sides of the initial interface. $M_{H}$ and $M_{L}$ are the molar masses of the heavy and light fluids, respectively. 

The dimensionless Navier-Stokes (NS) equations of compressible RT turbulence for a binary mixing fluid model are given by \cite{Gauthier2017,Luo2022,Luo2020}
\begin{equation}
	\frac{\partial \rho}{\partial {t}}+\frac{\partial (\rho u_{j})}{\partial x_{j}}=0,
	\label{ns1}
\end{equation}
\begin{equation}
	\frac{\partial (\rho u_{i})}{\partial t}+\frac{\partial (\rho u_{i} u_{j})}{\partial x_{j}}=-\frac{1}{Sr}\frac{\partial p}{\partial x_i}+\frac{1}{Re}\frac{\partial \sigma_{ij}}{\partial x_{j}}-\rho \delta_{i3},\label{ns2}
\end{equation}
\begin{equation}
	\begin{aligned}
		\frac{\partial\mathcal{E}}{\partial t}  +\frac{\partial (\mathcal{E} u_j )}{\partial x_{j}}&=-\frac{1}{{Sr}} \frac{\partial (p u_j)}{\partial x_{j}}+\frac{1}{{Re}} \frac{\partial ( u_{i} \sigma_{i j})}{\partial x_{j}}-\rho u_{3} 
		+\frac{\Delta_{H, L}^{*}}{\left(\gamma_{r}-1\right) {SrReSc}} \frac{\partial}{\partial x_{j}}\left(\rho T \frac{\partial c}{\partial x_{j}}\right)
		\\&
		+\frac{\gamma_{r}}{\left(\gamma_{r}-1\right) {Sr} {Re} {Pr}} \frac{\partial}{\partial x_{j}}\left(\frac{\partial T}{\partial x_{j}}\right),
	\end{aligned}
\end{equation}
\begin{equation}
	\frac{\partial (\rho c)}{\partial t}+\frac{\partial (\rho c u_{j})}{\partial x_{j}}=\frac{1}{ReSc}\frac{\partial }{\partial x_j}\left(\rho \frac{\partial c}{\partial{x_{j}}}  \right),\label{ns4}
\end{equation}
\begin{equation}
	\frac{p}{\rho T}=\frac{1}{1-A_t^2}\left( 1+A_t-2A_tc  \right), \label{ns5}
\end{equation}
where $u_{i}$ is the velocity component, $ \rho = \rho_H + \rho_L$ is the density, $p= p_H + p_L$ is the pressure, $T = T_H =T_L$ is the temperature, and $c$ is the concentration of the heavy fluid. $\sigma_{i j}=2 \mu S_{i j}-\frac{2}{3} \mu \delta_{i j} S_{k k}$ is the viscous stress, where $S_{i j}=\frac{1}{2}\left(\partial u_i / \partial x_j+\partial u_j / \partial x_i\right) $ is the strain rate, and $\mu = \mu_H =\mu_L$ is the dynamic viscosity, which is a constant. \cite{Luo2020} \cite{Luo2020,Gauthier2017} $\mathcal{E}=\frac{\rho e}{\left(\gamma_{r}-1\right) Sr}+\frac{1}{2} \rho u_{i} u_{i}$ is the total energy per unit volume, where $e=C_v T$ is the internal energy and $C_v=c C_{v,H}+(1-c) C_{v,L}$ is the specific heat at a constant volume. $\gamma_r$ is a reference variable for the ratio of specific heat at a constant pressure to that at a constant volume. $\Delta _{H,L}^*=\gamma_{H} C_{v, H}-\gamma_{L} C_{v, L}$ is the dimensionless difference in the specific heat at constant pressure,\cite{Luo2020,Luo2021} where $\gamma_H=5/3$ and $\gamma_L=5/3$ are the ratios of specific heat for heavy and light fluids, respectively.

The five reference governing parameters, including the stratification parameter $Sr$, Reynolds number $Re$, Atwood number $A_t$, Schmidt number $Sc$, and Prandtl number $Pr$, are defined by \cite{Luo2022,Luo2021} 
\begin{equation}
	\begin{aligned}
		& Sr=\frac{gL_r}{RT_r/M_r}, \ Re=\frac{g^{1/2}L_r^{3/2}}{\mu_r /\rho_r}, \ A_t=\frac{M_H-M_L}{M_H+M_L}, \\
		& Pr=\frac{\gamma_r\mu_rC_{v,r}}{\kappa_r},  \ {\rm and} \  Sc=\frac{\mu_r}{\rho_r D_r}.
	\end{aligned} 
\end{equation}
$Sr$ is related to the strength of flow compressibility (expansion and compression motions and density stratification). $c_0=\sqrt{p_r/\rho_r}=\sqrt{RT_r/M_r}$ is the isothermal speed of sound. In our numerical studies, the stratification parameter $Sr$ is 1.0, Atwood number $A_t$ is 0.5, Reynolds number $Re$ is 10000 or 30000, Prandtl number $Pr$ is 0.7, and Schmidt number $Sc$ is 1.0. Three spatial averages of a variable $\varphi$ are defined as
\begin{equation}
	\langle \varphi \rangle_{xy}(z, t)= \frac{1}{L_{x}L_{y}} \int_0^{L_y} \int_0^{L_x} \varphi(x,y,z,t) dxdy, \label{ave1}  
\end{equation} 
\begin{equation}
	\langle \varphi \rangle_{m}(t)= \frac{1}{L_{x}L_{y} (H_B+H_S)} \int_{-H_S}^{H_B}\int_0^{L_y} \int_0^{L_x} \varphi(x,y,z,t) dxdydz, \label{ave2}  
\end{equation} 
\begin{equation}
	\langle \varphi \rangle(t)= \frac{1}{2L_{x}L_{y} L_z} \int_{-L_z}^{L_z}\int_0^{L_y} \int_0^{L_x} \varphi(x,y,z,t) dxdydz, \label{ave2}        
\end{equation} 
where $H_B$ and $H_S$ are, respectively, the bubble and spike heights based on the $5\%$ threshold values of mean concentration\cite{Cook2001,Luo2021}. $\langle \cdot \rangle_{xy}$ denotes the horizontal average value, $\langle \cdot \rangle_{m}$ denotes the volume average in the mixing region, and $\langle \cdot \rangle$ denotes the volume average in the computational domain.

A filtering methodology is applied to the NS equations to obtain the LES equations. The filtering operation is 
\begin{equation}
	\bar{f}(\mathbf{x})=\int f(\mathbf{x}-\mathbf{r}) G(\mathbf{r}, \mathbf{x}; \Delta) d \mathbf{r}, 
\end{equation}
where $f$ is the arbitrary variable, $\bar{f}(\mathbf{x})$ is the filtered physical quantity, $G$ is the filter kernel, and $\Delta$ is the filter width. 
The Favre filtering (mass-weighted filtering) $\tilde{f}=\overline{\rho f} / \bar{\rho}$ is introduced for convenience in compressible flows, \cite{Aluie2013,Wang2018,Zhao2018} where the overbar denotes ordinary filtering and the tilde denotes Favre filtering.

For the LES of compressible RT turbulence, the filtered governing equations can be written as\cite{Luo2023}
	\begin{equation}
		\frac{\partial \bar{\rho}}{\partial t}+\frac{\partial\left(\bar{\rho} \tilde{u}_{j}\right)}{\partial x_{j}}=0,
	\end{equation}
	\begin{equation}
		\frac{\partial\left(\bar{\rho} \tilde{u}_{i}\right)}{\partial t}+\frac{\partial\left(\bar{\rho} \tilde{u}_{i} \tilde{u}_{j}+\bar{p} \delta_{i j}/Sr\right)}{\partial x_{j}}-\frac{1}{R e} \frac{\partial \tilde{\sigma}_{i j}}{\partial x_{j}}+\bar{\rho} \delta_{i3} \\
		=-\frac{\partial \tau_{i j}}{\partial x_{i}}+\frac{1}{Re} \frac{\partial\left(\bar{\sigma}_{i j}-\tilde{\sigma}_{i j}\right)}{\partial x_{i}}, \label{les.u}
	\end{equation}
	
	\begin{equation}
		\begin{aligned}
			\frac{\partial \tilde{\mathcal{E}}}{\partial t}+\frac{\partial}{\partial x_{j}} \left[\left(\tilde{\mathcal{E}}+\frac{\bar{p}}{Sr}\right) \tilde{u}_{j}\right] 
			-\frac{1}{{Re}} & \frac{\partial\left(\tilde{\sigma}_{i j} \tilde{u}_{i}\right)}{\partial x_{j}} 
			-\frac{\gamma_{r}}{\left(\gamma_{r}-1\right) Sr Re Pr} \frac{\partial}{\partial x_{j}}\left(\frac{\partial \tilde{T}}{\partial x_{j}}\right) \\
			&-\frac{\Delta_{H, L}^{*}}{\left(\gamma_{r}-1\right) SrReSc} \frac{\partial}{\partial x_{j}}\left(\bar{\rho} \tilde{T} \frac{\partial \tilde{c}}{\partial x_{j}}\right) 
			+ \bar{\rho} \tilde{u}_{3} = R_{\mathcal{E}}, \label{les.e}
		\end{aligned}
	\end{equation}
	
	\begin{equation}
		\frac{\partial\left(\bar{\rho} \tilde{c}\right)}{\partial t}+\frac{\partial \left(\bar{\rho} \tilde{c} \tilde{u}_{j}\right)}{\partial x_{j}} -\frac{1}{Re Sc} \frac{\partial}{\partial x_{j}} \left(\bar{\rho} \frac{\partial \tilde{c}}{\partial x_{j}}\right)=-\frac{\partial}{\partial x_{j}}\left[\bar{\rho} \left(\widetilde{c u_{j}} -\tilde{c} \tilde{u}_{j}\right) \right], \label{les.c}
	\end{equation}
	
	\begin{equation}
		\bar{p}-\frac{1+A_t-2A_t\tilde{c}}{1-A_t^2} \bar{\rho} \tilde{T} =
		-\frac{2A_t}{1-A_t^2}\left[\bar{\rho} \left(\widetilde{c T}-\tilde{c}\tilde{T}\right)\right],\label{les.p}
	\end{equation}
where 
\begin{equation}
	\begin{aligned}
		R_{\mathcal{E}} &= -\tilde{u}_{i} \frac{\partial \tau_{i j} }{\partial x_{j}}    
		-\frac{1}{\left(\gamma_{r}-1\right) Sr} \frac{\partial \bar{\rho}\left(\widetilde{u_{j} e}
			-\tilde{u}_{j} \tilde{e}\right) }{\partial x_{j}}  
		-\frac{\overline{p S_{k k}}-\bar{p}\tilde{S}_{kk}}{Sr}   		
		+\frac{\overline{\sigma_{ji} S_{ij}}-\tilde{\sigma}_{ji}\tilde{S}_{i j}}{Re}
		+\frac{\tilde{u}_{i}}{Re} \frac{\partial \left(\bar{\sigma}_{i j}-\tilde{\sigma}_{i j}\right) }{\partial x_{j}}  \\
		&+\frac{\gamma_{r}}{\left(\gamma_{r}-1\right) Sr Re Pr} \frac{\partial}{\partial x_{j}}\frac{\partial \left( \bar{T}-\tilde{T} \right)}{\partial x_{j}} 
		+\frac{\Delta_{H, L}^{*}}{\left(\gamma_{r}-1\right) SrReSc} \frac{\partial}{\partial x_{j}}\left[\bar{\rho}\left(\widetilde{T\frac{\partial c}{\partial x_j}}-\tilde{T} \frac{\partial \tilde{c}}{\partial x_{j}}\right)\right], 
	\end{aligned}
\end{equation}
and $\mathcal{E} = \frac{\bar\rho \tilde e}{\left(\gamma_{r}-1\right) Sr}+\frac{1}{2} \bar\rho \tilde{u}_{i} \tilde{u}_{i}$ is the resolved total energy, $\tilde\sigma_{i j}=2 \mu \tilde S_{i j}-\frac{2}{3} \mu \delta_{i j} \tilde S_{k k}$, and $\tilde S_{i j}=\frac{1}{2}\left(\partial \tilde{u}_{i} / \partial x_j+\partial \tilde{u}_{j} / \partial x_i\right)$.

There are many unclosed terms on the right-hand side of the LES equations. We only model the SGS stress $\tau_{i j}=\bar{\rho}\left(\widetilde{u_i u_j}-\tilde{u}_i \tilde{u}_j\right) $  and SGS diffusion $\psi_i=\bar{\rho} \left(\widetilde{c u_{i}} -\tilde{c} \tilde{u}_{i}\right) $ in LES. Other unclosed terms, including $ \bar{\rho}\left(\widetilde{u_{j} e}-\tilde{u}_{j} \tilde{e}\right) $, $\overline{p S_{k k}}-\bar{p}\tilde{S}_{kk} $, $\overline{\sigma_{ji} S_{ij}}-\tilde{\sigma}_{ji}\tilde{S}_{i j} $, $\bar{\sigma}_{i j}-\tilde{\sigma}_{i j} $, $\bar{T}-\tilde{T}$, and $\bar{\rho}\left(\widetilde{T\frac{\partial c}{\partial x_j}}-\tilde{T} \frac{\partial \tilde{c}}{\partial x_{j}}\right) $, are neglected and assumed to be 0 \cite{Bin2021,Mellado2005,ZhouH2021,Xie2018}.
In our current work, we adopt the top-hat filter, and the filter in one dimension can be written as \cite{Martin2000} 
\begin{equation}  
	\bar{f}_{i}=\frac{1}{4 n}\left(f_{i-n}+2 \sum_{j=i-n+1}^{i+n-1} f_{j}+f_{i+n}\right),
\end{equation}
where the filter width is $\Delta=2 n \Delta_{x}$, $\Delta_{x}$ is the grid size.

In the current work, the SGS stress $\tau_{i j}=\bar{\rho}\left(\widetilde{u_i u_j}-\tilde{u}_i \tilde{u}_j\right) $  and SGS diffusion $\psi_i=\bar{\rho} \left(\widetilde{c u_{i}} -\tilde{c} \tilde{u}_{i}\right) $ need to be modeled by the resolved variables to close the LES equations. We consider two SGS models: the velocity gradient model (VGM) and the dynamic Smagorinsky model (DSM).

The velocity gradient model (VGM) is a typical structural SGS model. \cite{Clark1979} The SGS stress $\tau_{i j}=\bar{\rho}\left(\widetilde{u_i u_j}-\tilde{u}_i \tilde{u}_j\right) $ and SGS diffusion $\psi_i$ are modeled as \cite{Clark1979,Luo2023}
\begin{equation}
	\tau_{i j}^{mod}=\frac{\Delta^{2}}{12}\bar{\rho} \frac{\partial \tilde{u}_{i}}{\partial x_{k}} \frac{\partial \tilde{u}_{j}}{\partial x_{k}},  \ 
	\psi_i^{mod} =\frac{\Delta^{2}}{12}\bar{\rho} \frac{\partial \tilde{u}_i}{\partial x_k} \frac{\partial \tilde{c}}{\partial x_k} . 	
\end{equation}

The Smagorinsky model is one of the most well-known functional SGS models.\cite{Smagorinsky1963}
For the compressible flows, the SGS stress can be written as \cite{Smagorinsky1963,Moin1991,Lilly1992,Garnier2009,Xie2018,Luo2023}
\begin{equation}
	\begin{gathered}
		\tau_{i j}^{A,mod}=\tau_{i j}^{mod}- \frac{\delta_{i j}}{3} \tau_{k k}^{mod}=-2 C_s^2 \bar{\rho} \Delta^2|\tilde{S}| ( \tilde{S}_{i j} - \frac{\delta_{i j}}{3} \tilde{S}_{kk} ),\label{DSM1}   \\
		\tau_{k k}^{mod} = 2 C_I \bar{\rho} \Delta^2|\tilde{S}|.
	\end{gathered}
\end{equation}
where $|\tilde S|$ is the characteristic filtered strain rate. The coefficients $C_s^2$ and $C_I$ can be dynamically determined by the Germano identity \cite{Germano1992} and least squares algorithm, giving rise to the dynamic Smagorinsky model (DSM), as \cite{Moin1991,Lilly1992,Xie2018,Luo2023}
\begin{equation}
	C_s^2=\frac{\left\langle\mathcal{L}_{i j} \mathcal{M}_{i j}\right\rangle_{xy}}{\left\langle\mathcal{M}_{k l} \mathcal{M}_{k l}\right\rangle_{xy}}, 
	C_I=\frac{\left\langle\mathcal{L}_{k k} (\beta - \hat{\alpha}) \right\rangle_{xy} }{\left\langle (\beta - \hat{\alpha})^2 \right\rangle_{xy}},
\end{equation}
where $\alpha_{ij}=-2\Delta^{2}\bar{\rho}|\tilde{S}|(\tilde{S}_{ij}
-\frac{\delta_{ij}}{3}\tilde{S}_{kk})$, 
$\beta_{ij}=-2\hat{\Delta}^2\hat{\bar{\rho}}|\hat{\tilde{S}}|(\hat{\tilde{S}}_{ij}
-\delta_{ij}\hat{\tilde{S}}_{kk}/3)$, 
$\mathcal{L}_{i j} = \hat{\bar{\rho}}(\widehat{\tilde{u}_{i}\tilde{u}_{j}}-\hat{\tilde{u}}_{i}\hat{\tilde{u}}_{j}) $, $\mathcal{M}_{ij}=\beta_{ij}-\hat{\alpha}_{ij}$, $\alpha=2\bar{\rho}\Delta^{2}{|\tilde{S}|}^{2}$ and  $\beta=2\hat{\Delta}^2\hat{\bar{\rho}}|\hat{\tilde{S}}|^2$. Here, an overbar denotes the filtering at scale $\Delta$, a tilde denotes the Favre filtering at scale $\Delta$, and a hat
represents a coarser filtering $\hat{\Delta} = 2\Delta$. 
The SGS diffusion $\psi_i$ is modeled as \cite{Smagorinsky1963,Moin1991,Lilly1992,Xie2018}
\begin{equation}
	\psi_{i}^{mod}=-C_d\Delta^{2}\bar{\rho}|\tilde{S}|\frac{\partial \tilde{c}}{\partial x_{i}},\
	C_d=\frac{\langle\mathcal{K}_{j}\mathcal{T}_{j}\rangle_{xy}}{\langle\mathcal{T}_{k}\mathcal{T}_{k}\rangle_{xy}}, \label{DSM2} 
\end{equation}
where $\mathcal{T}_{j}=-\hat{\Delta}^{2}\hat{\bar{\rho}}|\hat{\tilde{S}}|\frac{\partial \hat{\tilde{c}}}{\partial x_{j}}+\Delta^{2}\widehat{\bar{\rho}|\tilde{S}|\frac{\partial \tilde{c}}{\partial x_{j}}}$ and $\mathcal{K}_{j} = \hat{\bar{\rho}}(\widehat{\tilde{c}\tilde{u}_{j}}-\hat{\tilde{c}}\hat{\tilde{u}}_{j})$. The coefficients $C_s^2$, $C_I$, and $C_d$ are functions of height $z$. For numerical stability, if $C_s^2$, $C_I$, or $C_d$ is negative, then they are all taken as 0.

The numerical simulations of compressible RT turbulence are performed in a rectangular box of $1^2 \times 2$ ($0 \leq x \leq 1, 0 \leq y \leq 1, -1 \leq z \leq 1$) with periodic boundary conditions in the horizontal directions. In the vertical directions, no-slip boundary conditions of the velocity are used, gradient-free boundary conditions for concentration are used, and the temperature is fixed. \cite{Luo2020,Gauthier2017,Zhao2022} The initial density $ \rho_0$ for heavy or light fluid is an exponential distribution:
\begin{equation}
	\begin{aligned}
		\rho_0 &=(1 + A_t) \exp \left[-Sr(1+A_t) (x_{3}-z_0)\right]H_+(x_{3}-z_0) \\
		&+ (1 - A_t) \exp \left[-Sr(1-A_t)(x_{3}-z_0)\right]H_-(x_{3}-z_0),
	\end{aligned}
\end{equation}
where $z_0$ is the initial perturbation displacement, which is the superposition of sine and cosine functions with different wavenumbers.\cite{Dimonte2004pof,Luo2022,Youngs2013} The range of perturbation wavenumber is $5 \leq k=\sqrt{k_x^2+k_y^2} \leq 10$ in current work.
The regularized Heaviside functions $H_{\pm}(z)=[1\pm {\rm erf}(z/\delta)]/2$ is used to  smooth the flow field at the interface between two fluids, where $\operatorname{erf}(x)=\frac{2}{\sqrt{\pi}} \int_{0}^{x} e^{-s^{2}} ds$, and $\delta=4 \Delta_{x}$ is the pseudo-interface.\cite{Gauthier2017,Luo2022,Luo2020} 

For numerical simulations, an eighth-order central compact finite difference scheme is applied on a uniform grid to discretize the convective terms and SGS terms, and an eighth-order explicit central finite difference scheme is used to discretize the viscous terms. \cite{Lele1992,Wang2010} The diffusive terms are evaluated by the repeated application of first derivatives.
A third-order Runge-Kutta scheme is used for time advancing. \cite{Shu1988} An eighth-order numerical hyper-viscosity model is adopted to ensure the stability of the algorithm. \cite{Wang2010} A sufficiently large wave-absorbing layer at the top and bottom of the computational domain is used to mitigate the effects of acoustic waves. \cite{Reckinger2016,Luo2020}

\section{\label{sec:FNO}The Fourier neural operator}

The Fourier neural operator (FNO) aims to train on a finite set of input-output pairs to achieve mapping between two infinite-dimensional spaces. Compared with other neural network methods, the FNO method exhibits strong adaptability and generalization ability in dealing with high-dimensional and large-scale data. \cite{Li2022,guibas2021adaptive,rashid2022learning,pathak2022fourcastnet,li2022fouriergeo} The main idea of FNO is to utilize Fourier transformation to map high-dimensional data into the frequency domain and then use neural networks to learn the relationships between Fourier coefficients to approximate nonlinear operators. \cite{li2020fourier,Li2022}

\begin{figure*}[ht]\centering
	\setlength{\abovecaptionskip}{0.cm}
	\setlength{\belowcaptionskip}{-0.cm}
	\includegraphics[width=1.0\textwidth]{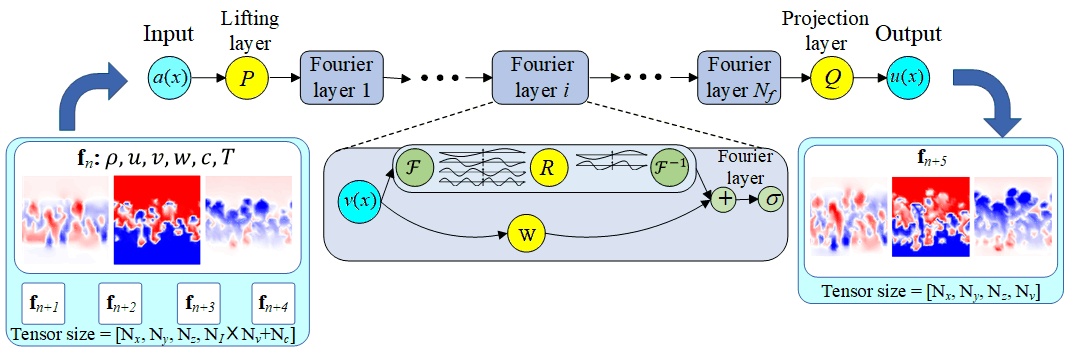} 
	\caption{The architecture of Fourier neural operator for 3D RT turbulence prediction. The middle is the FNO architecture, the lower left corner shows the input of the model, and the lower right corner shows the output. Here, $N_x=32, N_y=32$ and $N_z=64$ respectively represent grid resolutions in three coordinate directions, $N_v=6$ indicates the number of physical variables, $N_I=5$ is the number of time-nodes of input, and $N_c=3$ is the number of dimensions.}\label{fno.model}
\end{figure*}

The nonlinear mapping between two infinite-dimensional spaces is denoted as $ G^{\dagger}: \mathcal{A} \rightarrow \mathcal{U}$, where $\mathcal{A}=\mathcal{A} \left( D; \mathbb{R}^{d_a} \right)$ and $\mathcal{U}=\mathcal{U}\left(D; \mathbb{R}^{d_u}\right) $ respectively represent separable Banach spaces valued in $\mathbb{R}^{d_a}$ and $\mathbb{R}^{d_u}$, with $D \subset \mathbb{R}^d$ denoting a bounded open set, $\mathbb{R}$ representing the real number space. $\mathbb{R}^{d_a}$ and $\mathbb{R}^{d_u}$ are the value sets of input $a (x)$ and output $u (x)$, respectively, where $d_a$ and $d_u$ are the dimensions of $a (x)$ and $u (x)$, respectively. The Fourier neural operators learn an approximation of $G^{\dagger}$ by constructing a mapping $G:\mathcal{A} \times \Theta \rightarrow \mathcal{U}$ parameterized by $\theta \in \Theta$. Optimal parameters $\theta^{\dagger} \in \Theta$ can be determined through data-driven optimization methods, so that $G(\cdot, \theta^{\dagger}) \approx G^{\dagger}$.\cite{vapnik1999}

The Fourier neural operator employs iterative architectures $v_0\mapsto v_1\mapsto\ldots\mapsto v_{N_f}$, where $v_j$ for $j= 0,1, \ldots, N_f$ is a sequence of functions each taking values in $\mathbb{R}^{d_v}$. Here, $\mathbb{R}^{d_v}$ is the real numbers sets of $v_t(x)$, and $d_v$ is the dimension of $v_t(x)$.\cite{li2020b} The architecture of FNO is illustrated in Fig. \ref{fno.model}, comprising three main steps:\cite{li2020fourier,Li2022} 

\noindent (1) The input $a \in \mathcal{A}$ is lifted to a higher dimensional representation by a local transformation $P$, denoted as $v_0(x)=P\left( a(x)\right) $, parameterized by a shallow fully connected neural network.

\noindent (2) The high-dimensional $v_0(x)$ is updated through $N_f$ iterations, and each iteration is represented as
	\begin{equation}
		v_{t+1}(x)=\sigma\left(W v_t(x)+\left(\mathcal{K}(a; \phi) v_t\right)(x)\right), \ \forall x \in D,
		\label{eq9}
	\end{equation}
	where $\sigma: \mathbb{R} \rightarrow \mathbb{R}$ is a nonlinear activation function whose action is defined component-wise, $W: \mathbb{R}^{d_v} \rightarrow \mathbb{R}^{d_v}$ is a linear transformation, and $\mathcal{K}: \mathcal{A} \times \Theta_{\mathcal{K}} \rightarrow \mathcal{L}\left(\mathcal{U}\left(D ; \mathbb{R}^{d_v}\right), \mathcal{U}\left(D ; \mathbb{R}^{d_v}\right)\right) $ maps to bounded linear operators on $\mathcal{U}\left(D; \mathbb{R}^{d_v}\right) $ and is parameterized by $\phi \in \Theta_{\mathcal{K}}$.

\noindent (3) The output $u \in \mathcal{U}$ is the projection of $v_{N_f}(x)$ by the local transformation $Q: \mathbb{R}^{d_v} \rightarrow \mathbb{R}^{d_u}$, denoted as $u(x)=Q\left(v_{N_f}(x)\right)$, parameterized by a fully connected layer.

Replace the kernel integration operator in Eq. (\ref{eq9}) with a convolution operator defined in Fourier space. The Fourier transform and inverse transform of a function $f: D \rightarrow \mathbb{R}^{d_v}$ are respectively represented as $\mathcal{F}$ and $\mathcal{F}^{-1}$. The Fourier integral operator can be written as
\begin{equation}
		\left(\mathcal{K}(\phi) v_t\right)(x)=\mathcal{F}^{-1} \left\lbrace R_\phi \cdot \mathcal{F} \left\lbrace v_t  \right\rbrace \right\rbrace (x), \quad \forall x \in D,
		\label{eq10}
\end{equation}
where $R_\phi$ is the Fourier transform of a periodic function $\mathcal{K}: \bar{D} \rightarrow \mathbb{R}^{d_v \times d_v}$, parameterized by $\phi \in \Theta_{\mathcal{K}}$. $R_\phi$ is parameterized as a complex tensor $({k_{\max } \times d_v \times d_v})$, which contains a collection of truncated Fourier modes $R_\phi \in \mathbb{C}^{k_{\max } \times d_v \times d_v}$. Here, $\mathbb{C}$ denotes the complex space. The domain $D$ is assumed to be discretized with $n \in \mathbb{N}$ points, where $v_t \in \mathbb{R}^{n \times d_v}$. The finite-dimensional parameterization by truncating the Fourier series at a maximal number of modes $k_{\max }=\left|Z_{k_{\max }}\right|=\mid\left\{k \in \mathbb{Z}^d:\left|k_j\right| \leq k_{\max, j}\right., \left.j=1, \ldots, d\right\} \mid$. $\mathcal{F}\left\lbrace v_t  \right\rbrace \in \mathbb{C}^{k_{\max } \times d_v}$ can be obtained by truncating the higher modes. Therefore, $ (R_\phi \cdot \mathcal{F} \left\lbrace v_t  \right\rbrace )_{k, l} = \sum_{j=1}^{d_v} R_{\phi k, l, j}\left(\mathcal{F} \left\lbrace v_t  \right\rbrace \right)_{k, j}$,  $k = 1, \ldots, k_{\max }$, and $ j=1, \ldots, d_v$.\cite{li2020fourier,Li2022}


\section{\label{sec:data}Dataset description and FNO model settings}

The previous section introduced the architecture of Fourier neural operator. In this section, we will introduce the input–output	pairs shown in the lower left and lower right corners in Fig. \ref{fno.model}, as well as the parameters used in the FNO model in 3D compressible RT turbulence.

When using neural network methods for flow field prediction, we need to train the model with a large amount of data. In this study, the data comes from numerical simulations of compressible RT turbulence in a $1^2 \times 2$ rectangular box with a grid scale of $\Delta_{x}=1/128$. Thus, there are a total of $128^2 \times 256$ grid points. The Reynolds number is $Re=10000$. The data in current numerical simulations with $128^2 \times 256$ grids is referred to as ground truth (GT) data. The numerical simulations with $128^2 \times 256$ grid points are not sufficient to reach the spectral resolution of the Nyquist limit at $Re=10000$\cite{Lele1992}. For the lower-order statistics considered in this study, the results of ground truth are close to the DNS results. The comparison between $512^2 \times 1024$, $256^2 \times 512$ and $128^2 \times 256$ grids is provided in Appendix \ref{App.512}. When training the model, we need to downsample the high-fidelity data into the coarse grid, selecting uniformly $32^2 \times 64$ grids of the flow field data. We refer to the flow field with the coarse grids as the filtered ground truth (fGT) flow field. The Atwood number and stratification parameter are $A_t=0.5$ and $Sr=1.0$, respectively. Since the central plane has only $32^2$ grids, the wave number range of the initial disturbance is $5 \leq k \leq 10$, ensuring sufficient resolution for the initial disturbance. 

We randomly generate 305 sets of different initial conditions for numerical simulations. 300 sets of data are used for training the FNO model, while the remaining independent 5 sets are used as initial fields and ground truth results in the \emph{a posteriori} tests. The data used for training and prediction are different, and all statistical results are ensemble averages of the 5 cases.

In each case, the total numerical simulation time is $3\tau$, where the dimensionless time $\tau=\sqrt{{L_r}/(A_tg)}$, and the time step is $\Delta t = 0.001 \tau $ in numerical simulations with $128^2 \times 256$ grids. Here, we only use data from the first two dimensionless time during the training of the FNO model, resulting in a total of 2001 time points for the filtered flow field data. 
The FNO model can directly predict the flow field over a large time interval\cite{li2020fourier,Li2022}. The time step used by the FNO model, $\Delta t_{\rm{FNO}}$, is much larger than that of numerical simulations with $128^2 \times 256$ grids. We test $\Delta t_{\rm{FNO}}= 20\Delta t$, $40\Delta t$, $60\Delta t$, $80\Delta t$. The results show that the prediction performance of the FNO model is the best for $\Delta t_{\rm{FNO}}=60\Delta t $ ($\Delta t_{\rm{FNO}}=0.06 \tau$) in the \emph{a posteriori} tests.
We infer that FNO models with large time intervals can filter small-scale structures with high frequency. If the time step of the FNO model is too large, the prediction might miss some flow field structures. If $\Delta t_{\rm{FNO}}$ is too small, the prediction gives rise to the situation with more high-frequency structures in the flow field. High frequency structures are more difficult to learn and lead to more instability in the \emph{a posteriori} tests.
Although the FNO model trained with a smaller $\Delta t_{\rm{FNO}}$ accurately predicts the flow field in the first few steps, it is difficult to obtain convergent long-term predictive results. Therefore, we actually use only $(300 \times 34)$ sets of data during the training of the FNO model. In the \emph{a posteriori} tests, we need to predict for a longer time, $3\tau $, to examine the generalization performance of the FNO model over a longer time.

\begin{figure*}[ht]\centering
	\setlength{\abovecaptionskip}{0.cm}
	\setlength{\belowcaptionskip}{-0.cm}
	\includegraphics[width=0.8\textwidth]{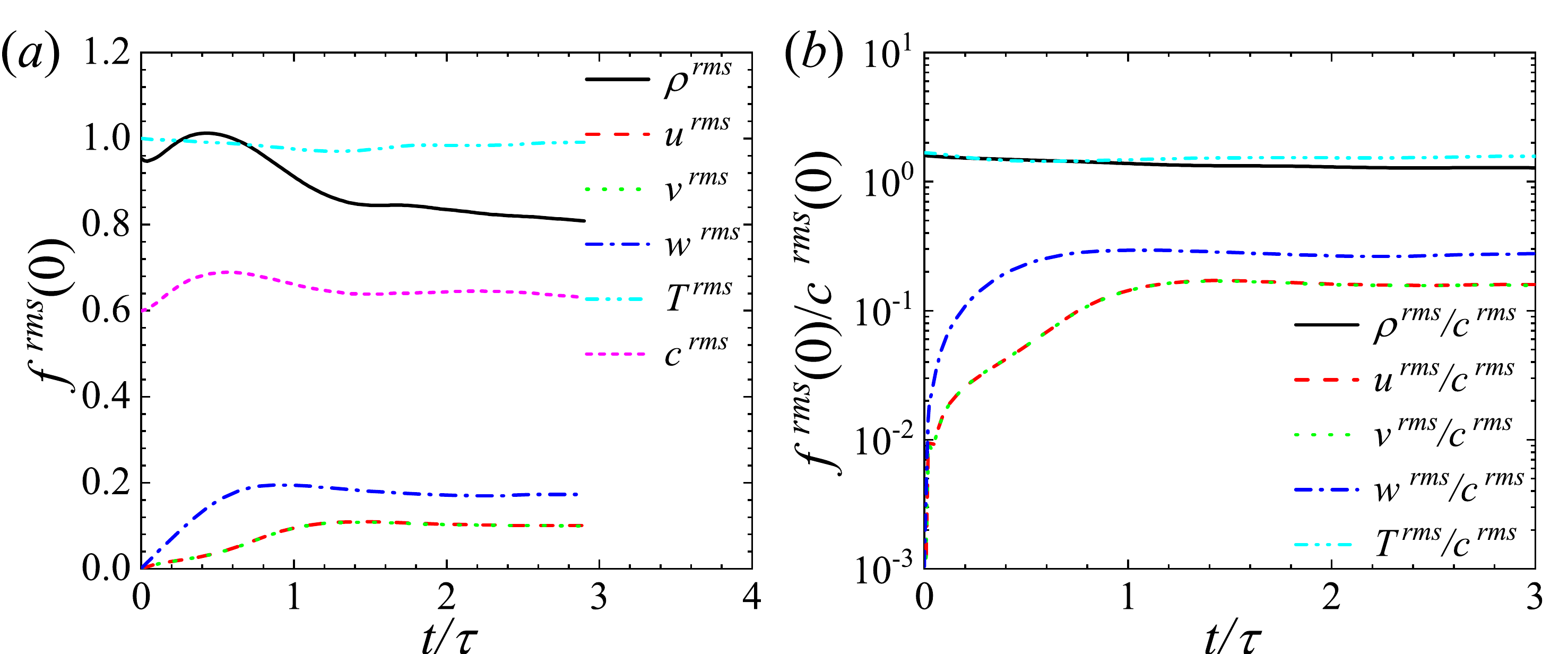} 
	\caption {Time evolution of the rms values of $\rho, u_{i}, T, c$ at $z=0$ in ground truth flow field:
		(a) $f^{rms}(0)$ and (b) $f^{rms}(0) / c^{rms}(0)$.}\label{fno-rms}
\end{figure*}
In previous studies of the FNO method, the flow was mostly incompressible, with only velocity fields, and the magnitudes of three velocity components were similar to each other. However, there are six physical quantities, including $\rho, u_{i}, T, c$, in current compressible RT turbulence. The magnitudes of six physical quantities are significantly different. Fig. \ref{fno-rms} shows the time evolution of the rms values of $\rho, u_{i}, T, c$ at $z=0$, which are ensemble averages of the ground truth data of 300 cases. The rms values of density and concentration increase first, then decrease, and eventually become nearly statistically steady over time, while the rms values of velocities increase first and then become nearly statistically steady. The rms value of temperature remains relatively steady. Additionally, there are large disparities in the magnitudes of the rms values of the six physical quantities, and the disparities change over time. For example, the ratio of the rms values of velocities and concentration increases by more than one order of magnitude over time. If we directly use the six physical fields for FNO model training, the FNO model would exhibit unstable performance in the \emph{a posteriori} tests. Therefore, we need to preprocess the training data. The dimensionless physical fields are defined as $f^*_{n} = f_{n} / f^{rms}_{n-1}(0)$, where $f$ represents any one of $\rho, u_{i}, T$ and $c$, as well as $f_{n}$ is the flow field at the $n$-th FNO time step, $f^*_{n}$ is the dimensionless flow field at the $n$-th FNO time step, and $f^{rms}_{n-1}(0) = \sqrt{\langle f_{n-1}(0)^2 \rangle }$ is the rms value at the center plane $z=0$ at the $(n-1)$-th FNO time step. The magnitudes of the six dimensionless physical quantities $f^*$ are consistent. It is important to note that for the data of the first and second FNO time steps, we use the rms values of the second time step for dimensionless transformation.

In summary, the training data for the FNO model consist of $[N_c \times N_t \times N_x \times N_y \times N_z \times N_v]$ samples, where $N_c = 300$ represents the number of independent cases, $N_t=34$ denotes the number of FNO time steps, $N_x=32, N_y=32, N_z=64$ represent the number of grid points in three coordinate directions, and $N_v=6$ indicates the number of physical variables. Therefore, there are a total of $[300 \times 34 \times 32 \times 32 \times 64 \times 6]$ samples, with 80\% of the data used for training and 20\% for testing. The dimensionless physical variables of the $n$-th FNO time step are denoted as $\mathbf{f}^*_{n}$. 
In the FNO model, the physical fields of the previous five FNO time steps $(\mathbf{f}^*_{n}, \mathbf{f}^*_{n+1}, \mathbf{f}^*_{n+2}, \mathbf{f}^*_{n+3}, \mathbf{f}^*_{n+4})$ are taken as input, which is similar to the situations of 3D incompressible HIT, free shear turbulence\cite{Li2022,Li2023}, and turbulent channel flows\cite{Wang2024}. If the number of history steps is small, the prediction results of the FNO model would be not so accurate. The physical variables at the sixth FNO time step $\mathbf{f}^*_{n+5,pre}$ are taken as output, as shown in Fig. \ref{fno.model}. Then, the flow fields of the 7-th time step $\mathbf{f}^*_{n+6,pre}$ are predicted using $(\mathbf{f}^*_{n+1}, \mathbf{f}^*_{n+2}, \mathbf{f}^*_{n+3}, \mathbf{f}^*_{n+4}, \mathbf{f}^*_{n+5,pre})$, and so on.
Therefore, we need to predict 46 steps to obtain data at $3\tau$ in the \emph{a posteriori} tests. During the training of the FNO model, we use the dimensionless physical fields $\mathbf{f}^*$. In the \emph{a posteriori} tests, we also obtain dimensionless physical fields, and then we need to use the rms values of the previous time step to obtain the true flow field ${f}_{n} = {f}^*_{n} \times f^{rms}_{n-1}(0)  $ at the current time step. 

\begin{table*}
	\caption{\label{FNO.parameters} Parameters of the FNO model}
	\centering
	\setlength\tabcolsep{10pt}
	\begin{tabular}{cccccc}
		\hline\hline
		Fourier layers & modes$(k_{\max, x }, k_{\max, y }, k_{\max, z })$ & width &  learning rate & batch size \\
		\hline
		4 & 10, 10, 20  & 99 &  0.001 & 10 \\
		\hline\hline
	\end{tabular}
\end{table*}
\begin{figure}[ht]\centering
	\setlength{\abovecaptionskip}{0.cm}
	\setlength{\belowcaptionskip}{-0.cm}
	\includegraphics[width=0.4\textwidth]{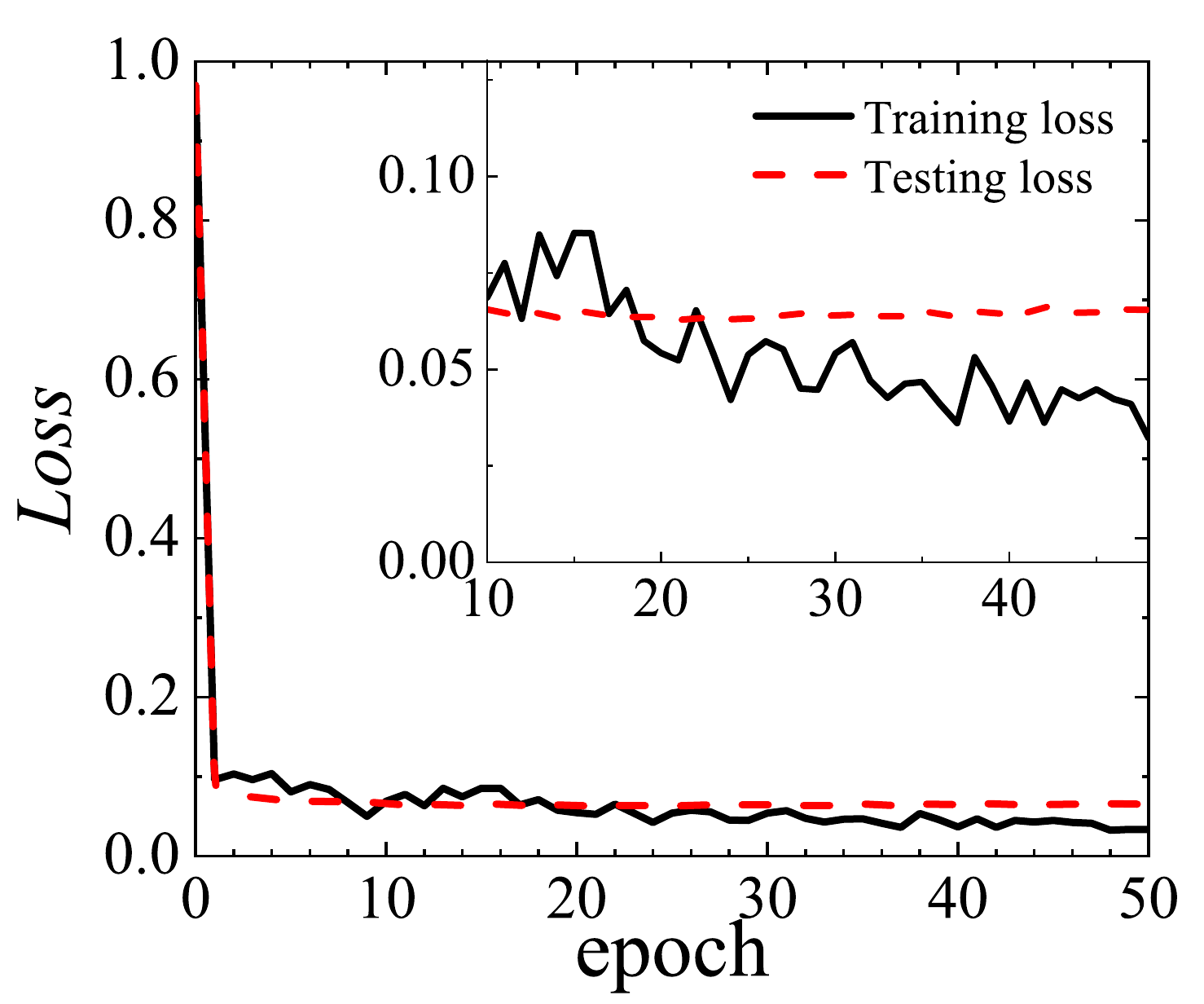} 	 	
	\caption {Learning curves of the FNO model for 3D RT turbulence.}\label{fig.loss}
\end{figure} 
A series of parameters need to be optimized in the FNO model. Table \ref{FNO.parameters} lists the parameters used in the FNO model. We choose 4 Fourier layers \cite{Li2022} with truncated Fourier modes $(k_{\max, x }=10, k_{\max, y }=10, k_{\max, z }=20)$, which are approximately 2/3 of the modes in Fourier space in each direction. The width is set to 99, the batch size is 10, and the learning rate is 0.001.\cite{li2020fourier,Li2022,Li2024} The ReLU function and adaptive moment optimization (Adam) algorithm are chosen as the activation function and optimizer, respectively. \cite{Kingma2014,Li2022} With these parameter settings, the FNO model achieves a good performance in the \emph{a posteriori} tests. The FNO model is trained for 50 epochs. Fig. \ref{fig.loss} shows the learning curves of the FNO model. The loss function is defined as
\begin{equation}
	Loss = \frac {\sqrt{\langle(\mathbf{f}^*-\mathbf{f}^*_{pre})^2\rangle}} {\sqrt{\left\langle \mathbf{f}^{*2}\right\rangle}}.
\end{equation}
It can be seen that the training loss consistently decreases, but the testing loss is nearly convergent after about 20 epochs.

\section{\label{sec:result} \emph{A POSTERIORI} TEST}

\begin{figure*}[ht]\centering
	\setlength{\abovecaptionskip}{0.cm}
	\setlength{\belowcaptionskip}{-0.cm}
	\includegraphics[width=0.9\textwidth]{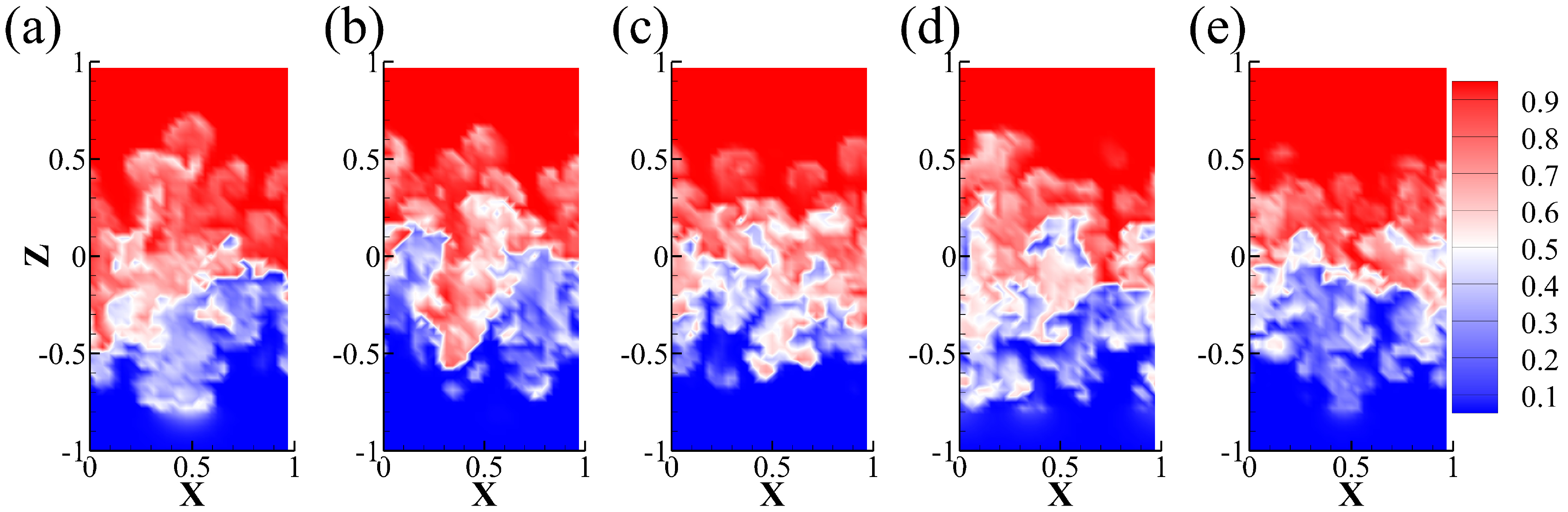}  
	\caption {The transient contours of the concentration $c$ of heavy fluid at $t/\tau =2.7 $ in ground truth flow field of 5 different cases.}\label{fig.con-fGT}
\end{figure*}

In this section, the \emph{a posteriori} tests for ILES and LES using the FNO, VGM, and DSM models of 3D compressible RT turbulence are performed. Fig. \ref{fig.con-fGT} shows the transient contours of the concentration of heavy fluid at $ t/\tau =2.7$ in ground truth flow field. We observe that the overall structures in different cases are similar to each other. Our focus is on predicting the statistics of the flow field and the spatial structures in the \emph{a posteriori} tests. All statistical results are ensemble averages of the 5 cases, which are independently generated with respective to the data used in the training of the FNO model. The first flow field obtained by the FNO model is at $t/\tau=0.3$. The flow field at $t/\tau \leq 0.24$ is the ground truth flow field.

\subsection{Model convergence and generalization on longer time}
	\begin{figure*}[ht]\centering
		\setlength{\abovecaptionskip}{0.cm}
		\setlength{\belowcaptionskip}{-0.cm}
		\includegraphics[width=0.8\textwidth]{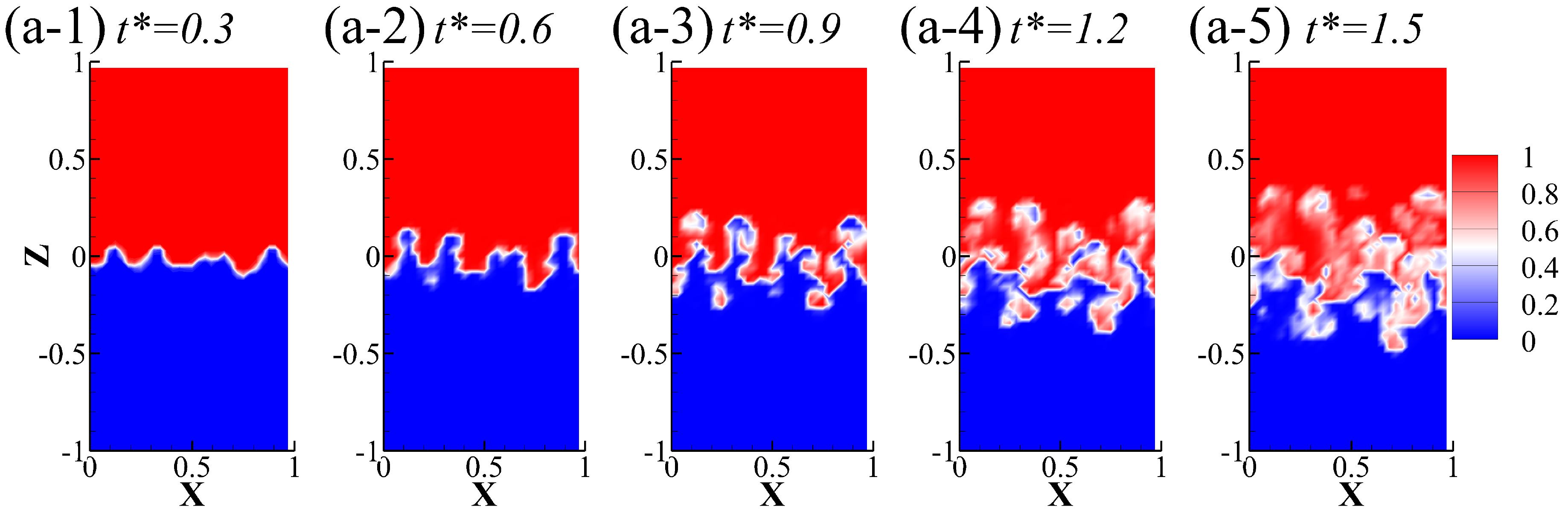} 
		\includegraphics[width=0.8\textwidth]{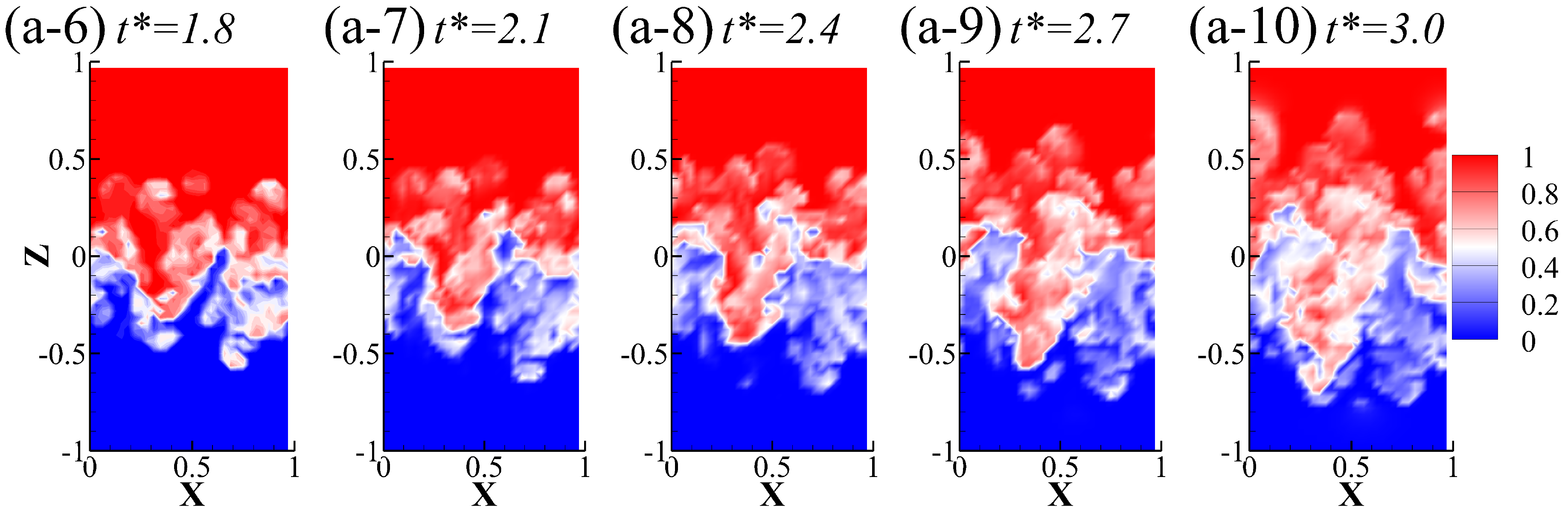} 
		\includegraphics[width=0.8\textwidth]{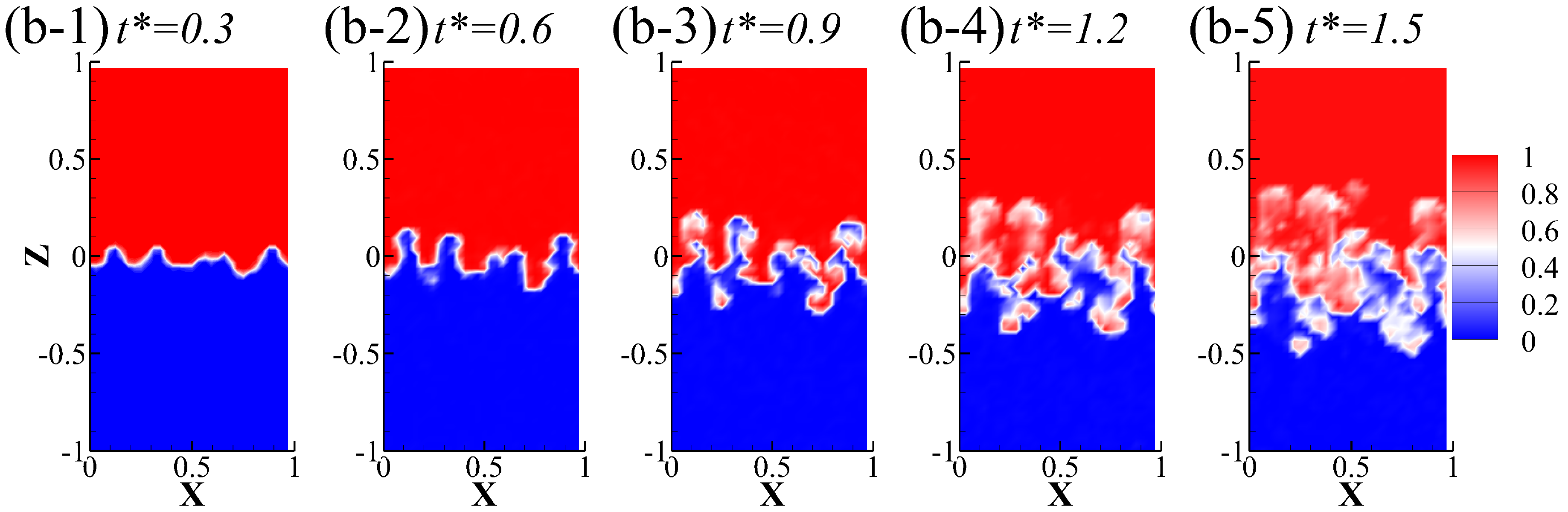} 
		\includegraphics[width=0.8\textwidth]{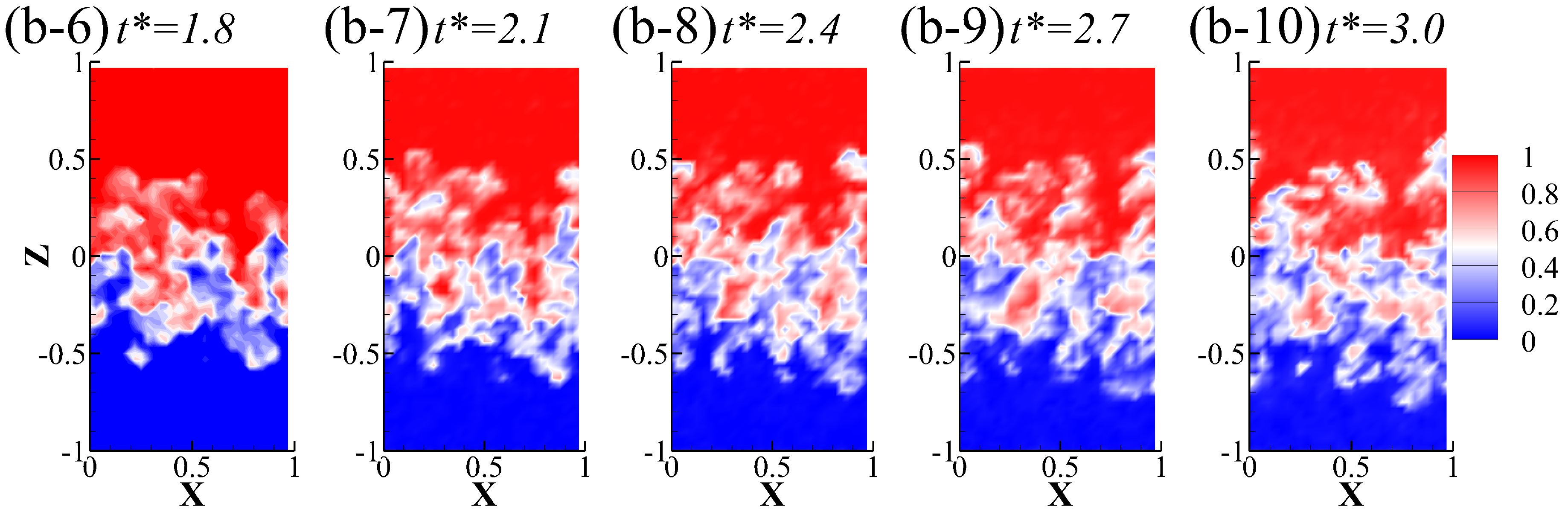} 	 	
		\caption {The contours of the concentration $c$ of heavy fluid at different times $t^*= t/\tau $: (a) filtered ground truth and (b) FNO.}\label{fig.con-c}
	\end{figure*} 
	The contours of the concentration $c$ of heavy fluid at different times $t^* = t/\tau$ in the numerical simulation with a grid resolution of $128^2 \times 256$ are presented in Fig. \ref{fig.con-c} (a). The total dimensionless time of the numerical simulation is $t^*=3$, and the contours depict the entire evolution process of simulated RT turbulence. In our simulations, the initial disturbance scale is relatively large, as shown in Fig. \ref{fig.con-c} (a–1). In the early stages ($t^* \leq 0.9$), disturbances grow independently, then bubbles and spikes compete or merge with each other. In the later stages, the RT flow gradually transitions into a turbulent state. Previous research results indicate that at the later stages of compressible RT turbulence, a certain degree of self-similarity has still been observed with the effect of asymmetric exponential initial density distribution. \cite{Luo2021} Here, we attempt to train the FNO model using data at $t^* \leq 2.0$. Moreover, when predicting the flow field using the FNO model, we can predict the entire development process, i.e., until $t^* = 3.0$. The data at time interval $2.0 < t^* \leq 3.0$ has never been used during training. Thus, we can evaluate the generalization of the FNO model for a longer time prediction.

	\begin{figure*}[ht]\centering
		\setlength{\abovecaptionskip}{0.cm}
		\setlength{\belowcaptionskip}{-0.cm}
		\includegraphics[width=0.8\textwidth]{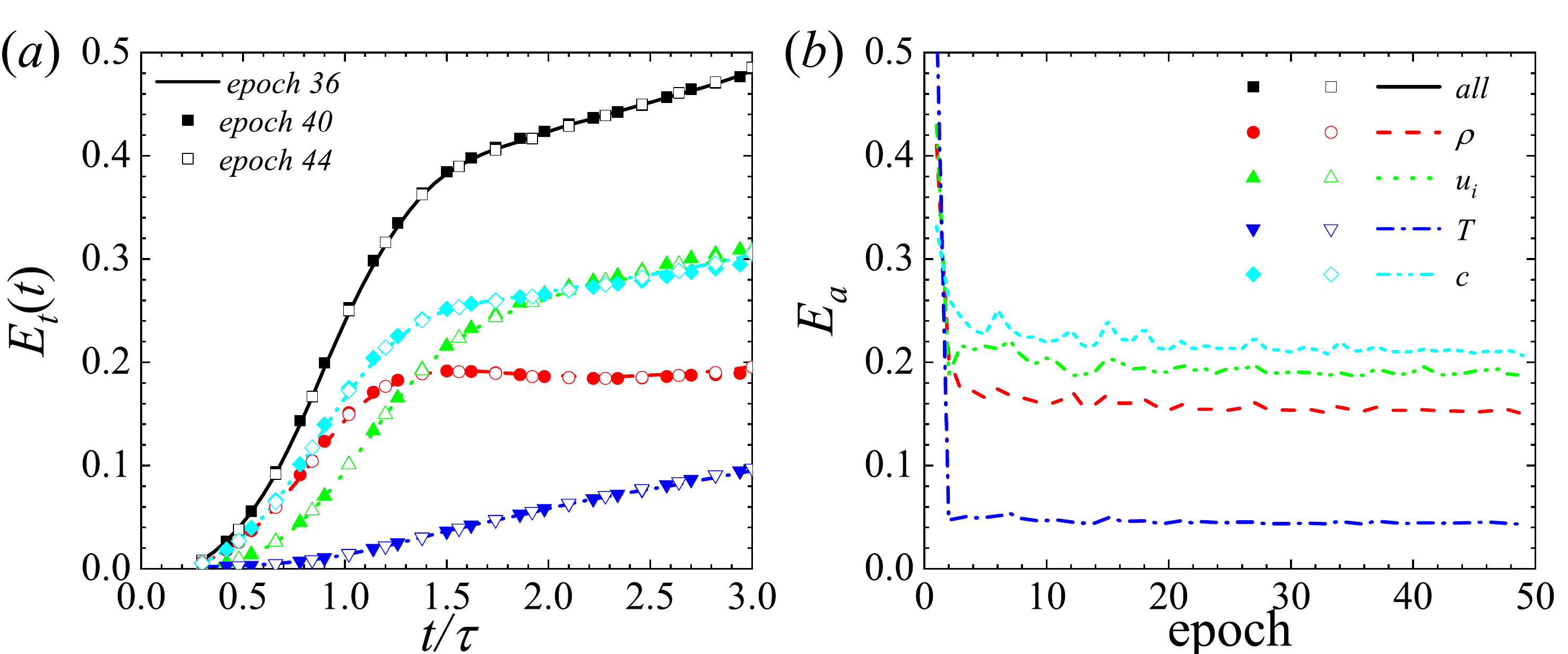} 	 	
		\caption {(a) Absolute error of models obtained with different training epochs  and (b) absolute error of different prediction steps.}\label{fig.error}
	\end{figure*} 
	During the training process of the FNO model, a model can be yielded at each epoch. Then, we first validate the convergence of the models obtained at different epochs. The absolute errors are defined as 	
	\begin{equation}
		E_t(t) = \sqrt{\left\langle\left(H-H^{pre}\right)^2\right\rangle}, \quad E_a = \frac{\sum_1^{N_t} E_t(t)}{N_t},
	\end{equation}	
	here, $H = \rho, u_i, T$, or $c$ represents the data from filtered ground truth (fGT) flow field, $H^{pre}$ represents the flow field predicted by the FNO model, and $N_t$ represents the number of time points. $E_t(t)$ denotes the error of the flow field predicted by the FNO model as the prediction steps increase (i.e., the time of RT turbulence development). The absolute errors $E_t(t)$ at epochs 36, 40, and 44 are depicted in Fig. \ref{fig.error} (a). $E_a$ is the average error of all prediction steps, and the variation of the absolute error $E_a$ of the predicted flow field by the FNO model with epochs is illustrated in Fig. \ref{fig.error} (b).	
	We observe that during the initial epochs, the absolute error of the model rapidly decreases first, followed by a period of fluctuation in the \emph{a posteriori} test. With an increase in training epochs, approximately after 30 epochs, the results predicted by the models in the \emph{a posteriori} test tend to converge. In Fig. \ref{fig.error} (a), we observe that the absolute errors $E_t(t)$ of the FNO models at three selected epochs overlap with each other. Although we only use data at $t^* \leq 2.0$ during training FNO model, the absolute error in prediction does not increase rapidly at $t^* \geq 2.0$.

	\begin{figure*}[ht]\centering
		\setlength{\abovecaptionskip}{0.cm}
		\setlength{\belowcaptionskip}{-0.cm}
		\includegraphics[width=0.8\textwidth]{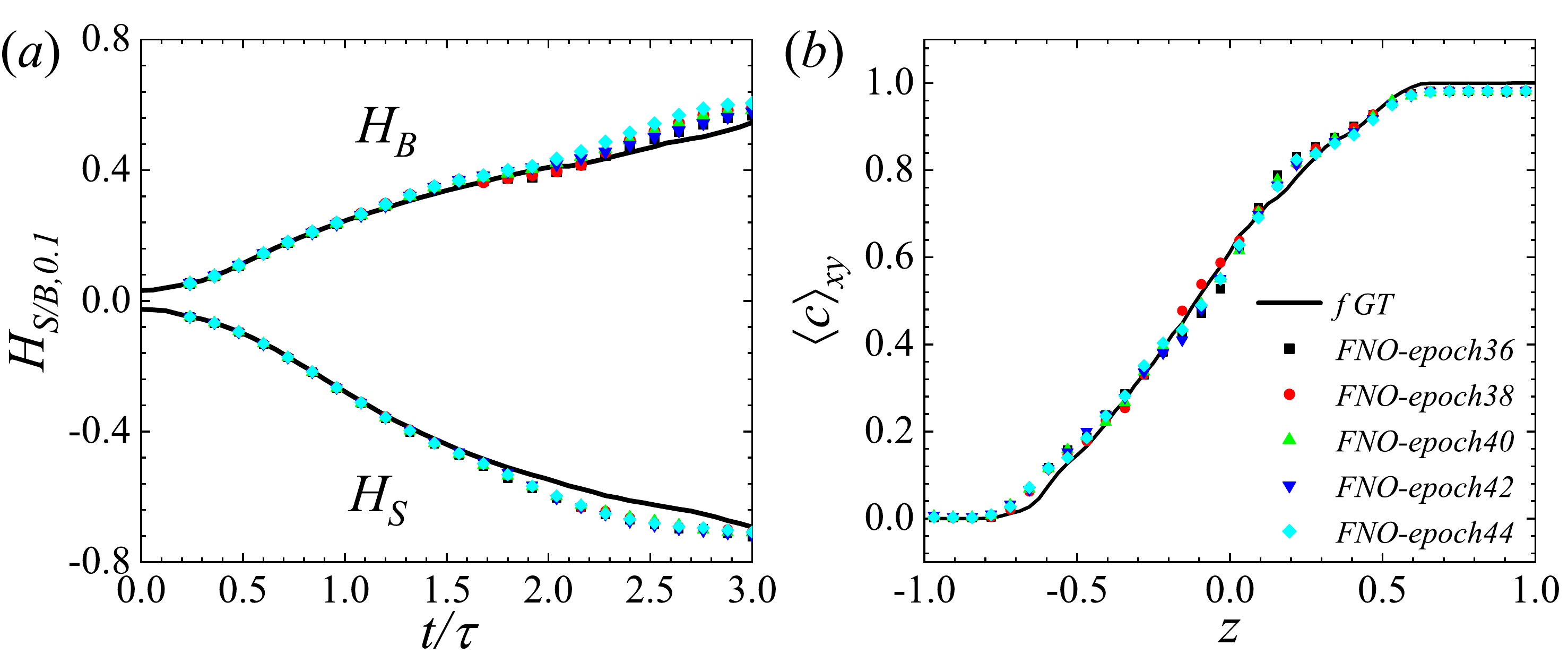} 	 	
		\caption {(a) The evolution of heights of bubble $H_{B,0.1}$ and spike $H_{S,0.1}$ and (b) the concentration profiles $\langle c \rangle_{xy}$ at $t/\tau =2.1 $ for different epochs of FNO.}\label{fig.H0.1}
	\end{figure*} 
	Fig. \ref{fig.H0.1} displays the predictions of the FNO models at epochs 36, 38, 40, 42, and 44, compared to the filtered ground truth (fGT): Fig. \ref{fig.H0.1} (a) shows the evolution of bubble and spike heights, where $H_{B,0.1}$ or $H_{S,0.1}$ represents the height of bubble or spike calculated based on a $10\%$ threshold of concentration field, where the mean concentration is 0.9 or 0.1;\cite{Cook2001,Livescu2009,Luo2022} Fig. \ref{fig.H0.1} (b) provides the concentration profiles $\langle c \rangle_{xy}$ at time $t/\tau = 2.1$. We observe that the FNO models can accurately predict the evolution of mixing heights over time and the change in average concentration with height very well as compared to the filtered ground truth (fGT) in the \emph{a posteriori} test. The prediction results of five FNO models overlap with each other, indicating the convergence of the FNO models. The statistics demonstrate that the FNO model trained with the data at $t^* \geq 2.0$ exhibits time generalization in predicting the concentration field and can be used for long-term predictions.
	
	In the compressible RT turbulence at current parameters, since the profile of the average concentration is nearly self-similar \cite{Luo2021}, the mixing heights predicted by the FNO models at $t^* \geq 2.0$ still match well with the filtered ground truth. Fig. \ref{fig.con-c} (b) presents the contours of the concentration of heavy fluid at different times $t^*$ predicted by the FNO model at epoch 40. 
	It can be observed that the contours of the concentration predicted by the FNO model are essentially the same as those from filtered ground truth in the early stages. But errors gradually accumulate over time. At the later stages, differences between the contours predicted by the FNO model and those from filtered ground truth emerge, although the basic flow structures remain similar. At $t^* \geq 2.0$, the flow field predicted by the FNO model is close to the result of solving NS equations. Qualitatively, the flow field does not exhibit any phenomena that are inconsistent with the ground truth results of RT turbulence.

	\begin{figure*}[ht]\centering
		\setlength{\abovecaptionskip}{0.cm}
		\setlength{\belowcaptionskip}{-0.cm}
		\includegraphics[width=0.8\textwidth]{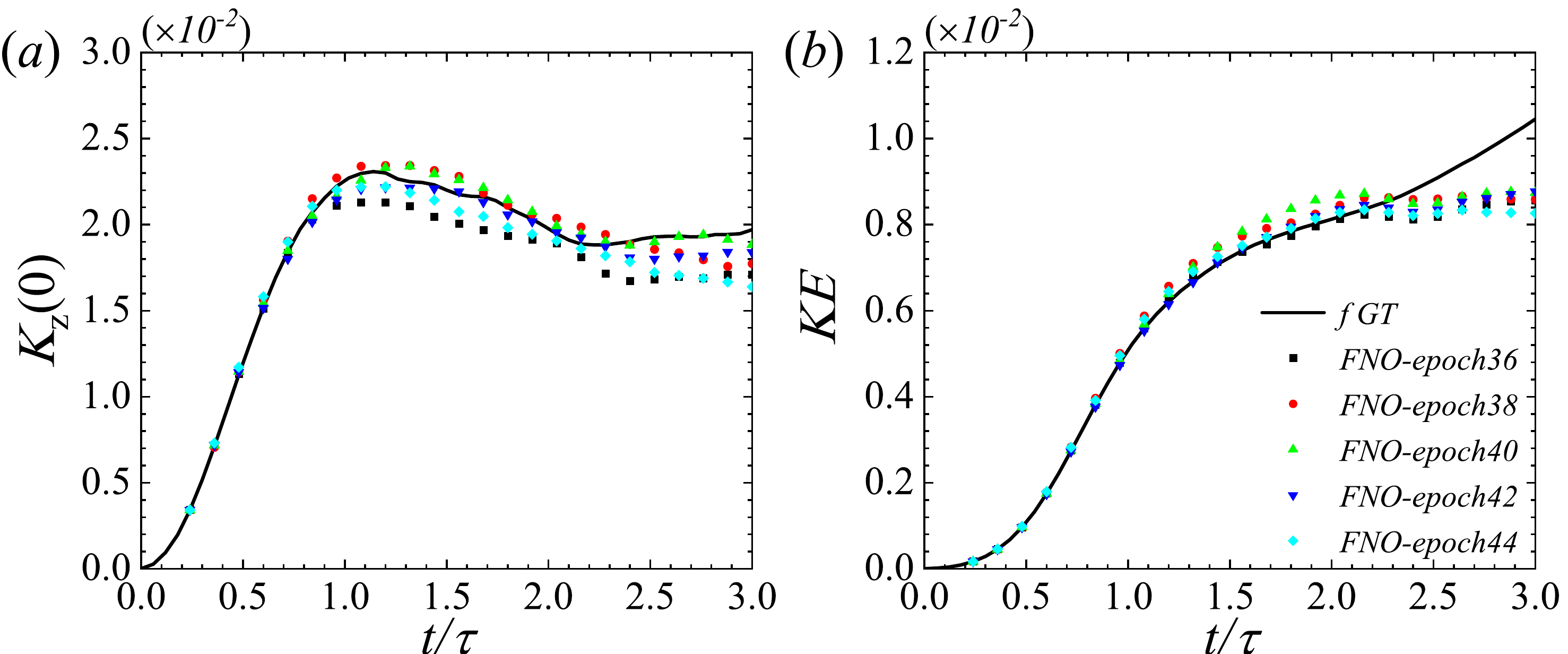} 	 	
		\caption {Time evolution of spatially averaged kinetic energy in the central $x-y$ plane, $K_z(0)$, and in the computational domain, $KE$, for different epochs in the FNO: (a) $K_z(0)$ and (b) $KE$.}\label{fig.KE-10000}
	\end{figure*} 
    Fig. \ref{fig.KE-10000} illustrates the time evolution of spatially averaged kinetic energy $K_z(0)= \langle \frac{\bar\rho}{2} \tilde{u}_i \tilde{u}_i \rangle_{xy}$ at $z=0$ plane and the kinetic energy $KE= \langle \frac{\bar\rho}{2} \tilde{u}_i \tilde{u}_i \rangle$ in the computational domain predicted by FNO models at epochs 36, 38, 40, 42, and 44 in the \emph{a posteriori} test.
    The kinetic energy predicted by the five models nearly overlaps with each other, confirming the convergence of the FNO models. The time evolution of the kinetic energy $K_z(0)$ at $z=0$ can be roughly divided into three stages: a rapid growth stage at $t^* \leq 1.0$, a gradual decrease stage at $1.0 < t^* \leq 2.0$, and a steady stage at $t^* > 2.0$. Surprisingly, the FNO models perform well in predicting at all three stages, even though the filtered ground truth (fGT) data at the third stage was not used during training. This suggests that the FNO model has learned certain physical principles at $z=0$. However, for averaged kinetic energy $KE$ in the computational domain, the FNO models predict results close to filtered ground truth up to $t^* \leq 2.2$, but deviate from filtered ground truth thereafter. This deviation may mainly be due to the asymmetric exponential initial density stratification, resulting in poorer self-similarity of the density field.
    The region outside the mixing layer is in a hydrostatic state, which makes it difficult to learn the evolution of the flow field at $t^* > 2.0$ using the FNO model trained with data $t^* \leq 2.0$.
    
    Although FNO is purely data-driven, it exhibits good generalization performance on longer-time predictions in the decaying HIT.\cite{Li2023} Here, we test generalization performance of FNO in RT turbulence and observe that the predictions are satisfactory. The generalization ability of the FNO model benefits from the self similarity in the later stage of RT turbulence. However, for compressible RT turbulence, self similarity exists within a certain parameter and time range.\cite{Luo2021,Luo2022,Gauthier2017} The generalization performance is good for the current parameters and time range. For the current statistical results, only kinetic energy (Fig. \ref{fig.KE-10000} (b)) has deviations at $t>2$, but this degree of deviation is acceptable.

\subsection{\label{sec:LES}Comparison between FNO model and traditional LES methods}
    In this section, we will compare the performance of the FNO model and the traditional SGS models in predicting compressible RT turbulence. In the \emph{a posteriori} test, the FNO model at epoch 40 is used. As a comparison, the LESs are also simulated starting from the 5-th FNO prediction step ($t_0=0.24 \tau $). The grids of LESs are consistent with the FNO model, $32^2 \times 64 $.

	\begin{figure*}[ht]\centering
		\setlength{\abovecaptionskip}{0.cm}
		\setlength{\belowcaptionskip}{-0.cm}
		\includegraphics[width=0.8\textwidth]{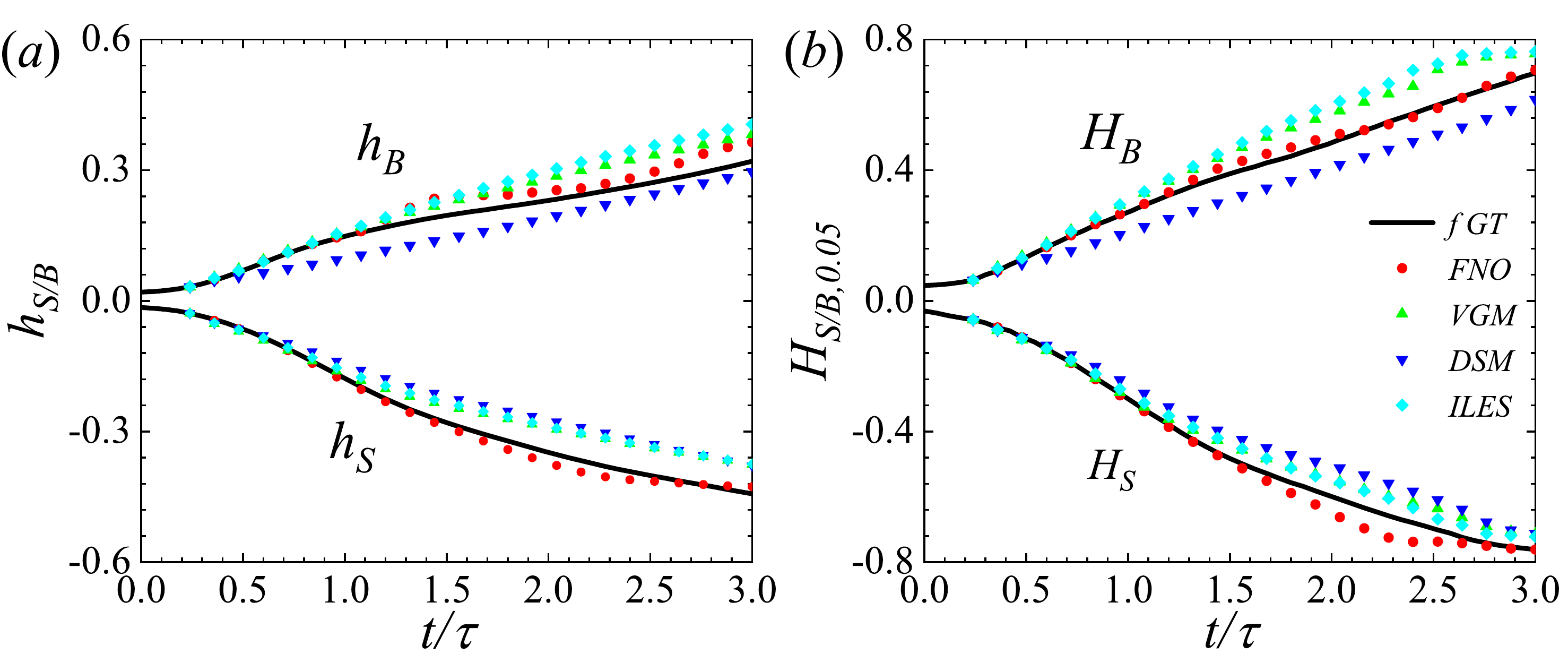} 	 	
		\caption {The evolution of bubble and spike heights: (a) $h_{B/S}$ and (b) $H_{B/S,0.05}$.}\label{fig.hbs}
	\end{figure*} 
	Fig. \ref{fig.hbs} illustrates the time evolution of bubble and spike heights from FNO model, different LES methods, and filtered ground truth (fGT). For two mixing heights, $h_{B/S}$ \cite{Gauthier2017,Dimonte2004pof,Livescu2010,Luo2022} and $H_{B/S,0.05}$\cite{Zhou2019pof,Livescu2010,Livescu2009}, and the mixing heights $H_{B/S,0.1}$ provided in Fig. \ref{fig.H0.1}, the results predicted by the FNO model have an overall closer agreement with the filtered ground truth throughout the entire evolution process. Consistent with the previous studies,\cite{Luo2023} the bubble heights predicted by the VGM model and ILES are higher than the filtered ground truth, while the bubble heights predicted by the DSM model are lower than the filtered ground truth. Different from previous studies, the spike heights predicted by the ILES, VGM, and DSM models are all lower than the filtered ground truth. In summary, the FNO model is significantly superior to traditional SGS models and ILES in predicting mixing heights.
	
	\begin{figure*}[ht]\centering
		\setlength{\abovecaptionskip}{0.cm}
		\setlength{\belowcaptionskip}{-0.cm}
		\includegraphics[width=0.8\textwidth]{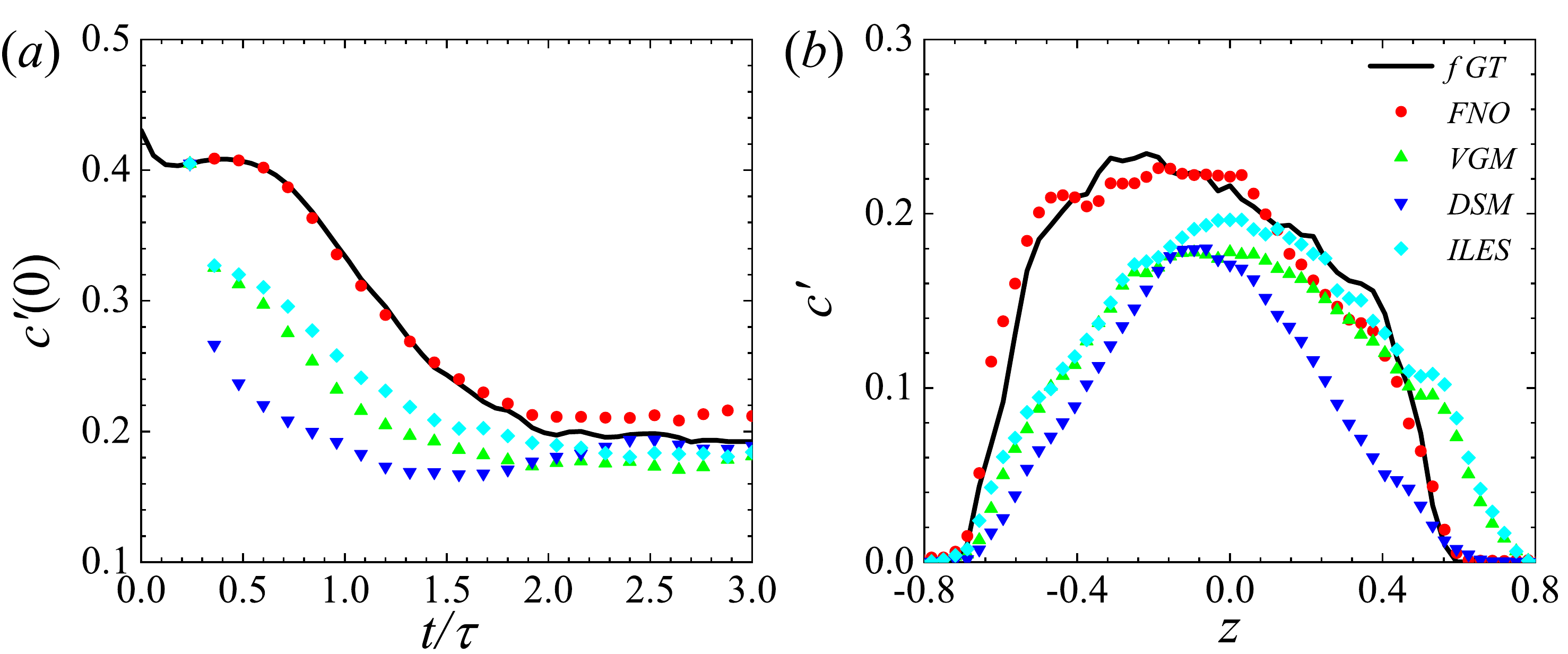} 	 	
		\caption {(a) Time evolution of the rms value of concentration fluctuation at $z=0$ and (b) profiles of the rms value of concentration fluctuation as functions of $z$ at $t/\tau =1.8$.}\label{fig.c-rms}
	\end{figure*} 
    For the average values of concentration, both the FNO model and traditional SGS models can predict relatively well. In Fig. \ref{fig.c-rms}, we display the time evolution of the root mean square (rms) values of concentration fluctuation $c'=\sqrt{\langle (c - \langle c \rangle_{xy})^2\rangle_{xy} } $ at $z=0$, and the profiles of the rms values of concentration fluctuation at $t/\tau =1.8$. The time evolution of $c'(0)$ at $z=0$ can be roughly divided into three stages: a steady stage at $t/\tau \leq 0.6$, a rapid decline stage at $0.6 < t/\tau \leq 2.0$, and a steady stage at $t/\tau > 2.0$. The FNO model provides excellent predictions for $c'(0) $ at all three stages, with a small deviation at $t/\tau \geq 1.9$. However, the results predicted by traditional SGS models are not satisfactory. Traditional SGS models significantly underestimate the rms values of concentration fluctuation at $z=0$ for the early stages of mixing ($t/\tau \leq 2.0$), gradually approaching the filtered ground truth as the development of RT turbulence. Moreover, Fig. \ref{fig.c-rms} (b) demonstrates that the FNO model gives the best prediction of the overall profiles of $c'$. In contrast, the shapes of the profiles predicted by traditional ILES, VGM, and DSM models deviate considerably from the filtered ground truth. Similar to the filtered ground truth, the profiles of $c'$ predicted by the FNO model are asymmetric, with the maximum value occurring at $z<0$, attributed to the smaller structure of spikes compared to bubbles and the more intense turbulent fluctuations in spike regions \cite{Livescu2009}. However, the profiles predicted by traditional LESs are generally more symmetric.
    
	\begin{figure*}[ht]\centering
		\setlength{\abovecaptionskip}{0.cm}
		\setlength{\belowcaptionskip}{-0.cm}
		\includegraphics[width=0.8\textwidth]{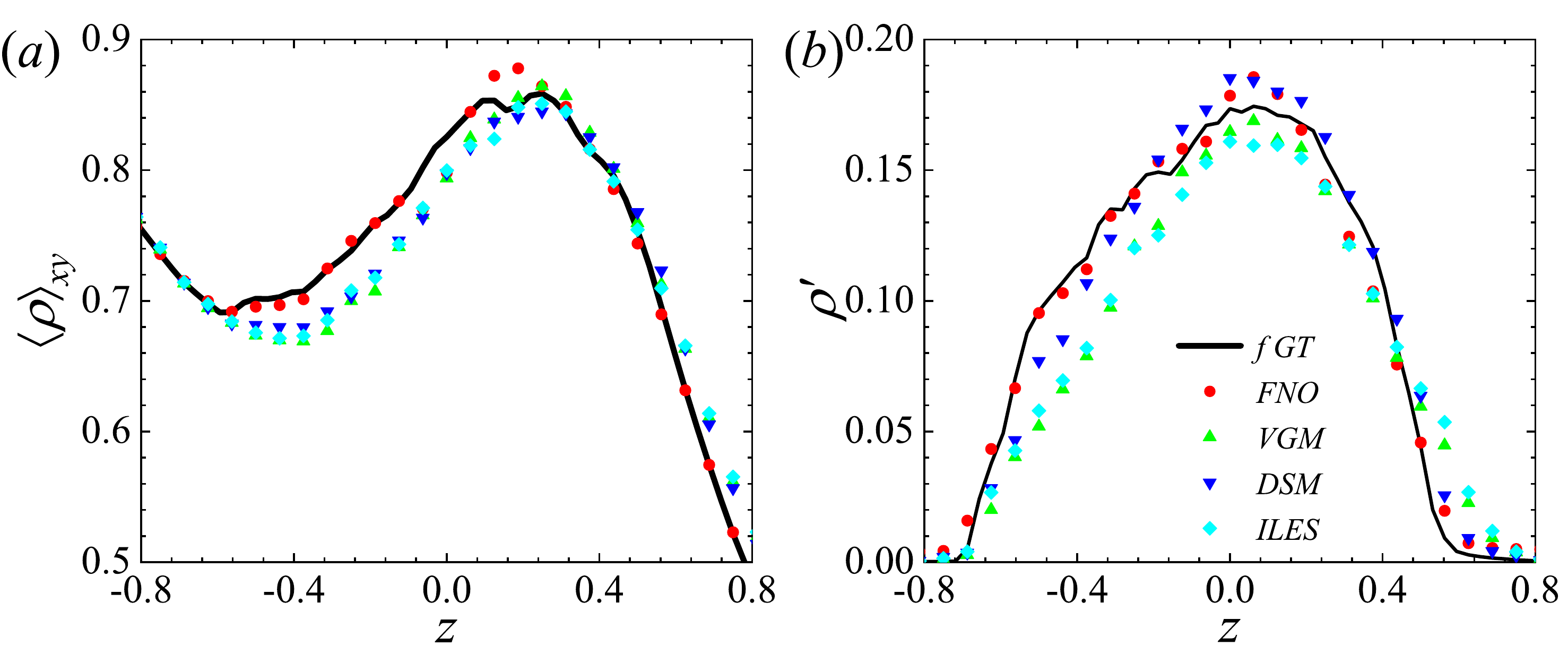} 	 	
		\caption {(a) Profiles of mean density $\langle \rho \rangle_{xy}$ and (b) profiles of the rms value of density fluctuation $\rho'$ as functions of $z$ at $t/\tau =1.8$.}\label{fig.rho-pro}
	\end{figure*} 
    Fig. \ref{fig.rho-pro} displays the profiles of the mean density $\langle \rho \rangle_{xy}$ and rms value of density fluctuation $\rho'$ at $t/\tau = 1.8$. It is shown that the results of the FNO model are the closest to the filtered ground truth. The profiles of $\langle \rho \rangle_{xy}$ predicted by the FNO model agree well with the filtered ground truth both within and outside the mixing region. The mean density predicted by SGS models and ILES is lower than the filtered ground truth in the mixing region and higher than the filtered ground truth outside the bubble region. The profiles of $\rho'$ are nearly asymmetric, and the predicted profiles by all models are nearly asymmetric as well, with the result predicted by the FNO model being the best among them.
	
	\begin{figure*}[ht]\centering
		\setlength{\abovecaptionskip}{0.cm}
		\setlength{\belowcaptionskip}{-0.cm}
		\includegraphics[width=0.8\textwidth]{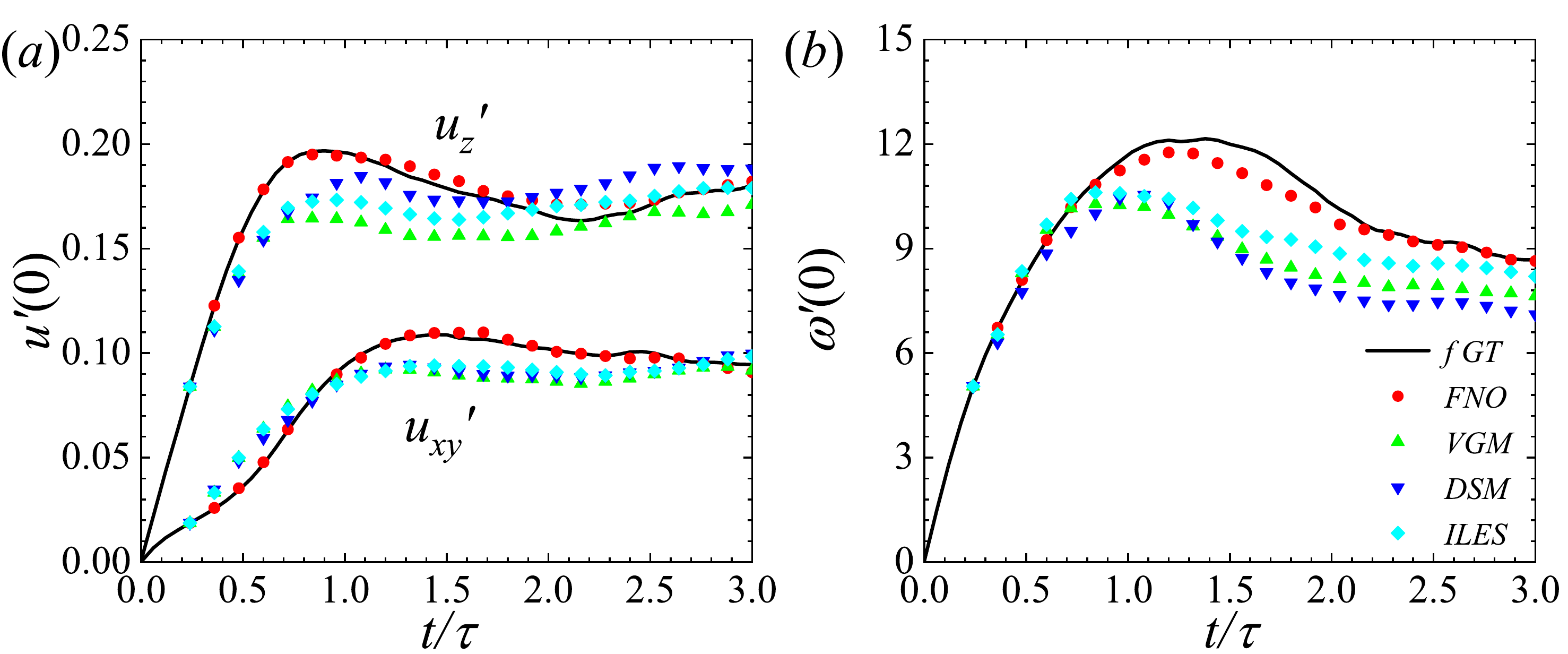} 	 	
		\caption {Time evolution of the rms values of (a) horizontal and vertical velocities, $u'_{xy}(0)$ and $u'_{z}(0)$, and (b) vorticity, $\omega'(0)$, at $z=0$.}\label{fig.uw-rms}
	\end{figure*} 
    \begin{figure*}[ht]\centering
		\setlength{\abovecaptionskip}{0.cm}
		\setlength{\belowcaptionskip}{-0.cm}
		\includegraphics[width=0.8\textwidth]{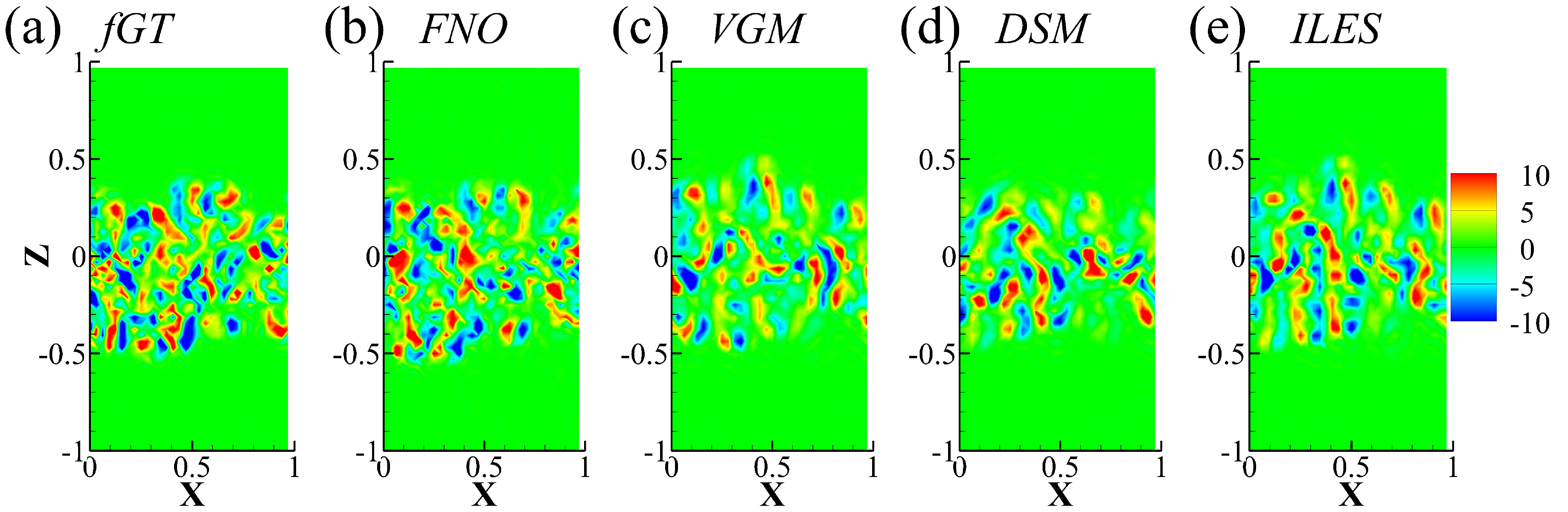} 	 	
		\caption {The contours of vorticity component $\omega_2$ at $t/\tau =1.5 $.}\label{fig.con-w2}
    \end{figure*} 
    Fig. \ref{fig.uw-rms} (a) illustrates the time evolution of the rms values of horizontal velocity $u'_{xy}(0)$ and vertical velocity $u'_z(0)$ at $z=0$. Both the rms values of vertical and horizontal velocities initially increase rapidly and then become nearly statistically steady over time. We observe that the rms value of vertical velocity reaches its maximum earlier, at approximately $0.8t/\tau$, while the rms value of horizontal velocity peaks around $1.3t/\tau$. In the later stages, the ratio of rms values between vertical and horizontal velocities becomes nearly statistically steady at around 1.6–1.8, which is close to 1.8 in 3D incompressible RT turbulence at low Atwood numbers \cite{Boffetta2010}, and higher than the results in 2D incompressible and compressible RT turbulence\cite{Zhou2013, Luo2021}. It can be observed that the FNO model perfectly predicts the time evolution of rms values of velocity components. The predictions of the ILES, VGM, and DSM models exhibit deviations in the growth stage, but the final results show no significant differences. The rms values of horizontal velocity predicted by three traditional LES models overlap substantially, but significant differences are observed in the rms values of vertical velocity.
    
    Fig. \ref{fig.uw-rms} (b) depicts the time evolution of the rms values of vorticity magnitude $\omega'(0)$ at $z=0$. The rms value of vorticity magnitude also initially increases, then slowly decreases, gradually becomes nearly steady. The time that $\omega'(0)$ reaches a maximum value closes to that of $u'_{xy}(0)$. The FNO model gives excellent predictions, while the three traditional LES models underestimate the rms value of vorticity magnitude. The results of ILES are better than those of the VGM and DSM models. Fig. \ref{fig.con-w2} shows the instantaneous vorticity component $\omega_2$ at $t/\tau = 1.5$. Compared to traditional LESs, the FNO model has a closer agreement with the vorticity structures of filtered ground truth.

	\begin{figure*}[ht]\centering
		\setlength{\abovecaptionskip}{0.cm}
		\setlength{\belowcaptionskip}{-0.cm}
		\includegraphics[width=0.8\textwidth]{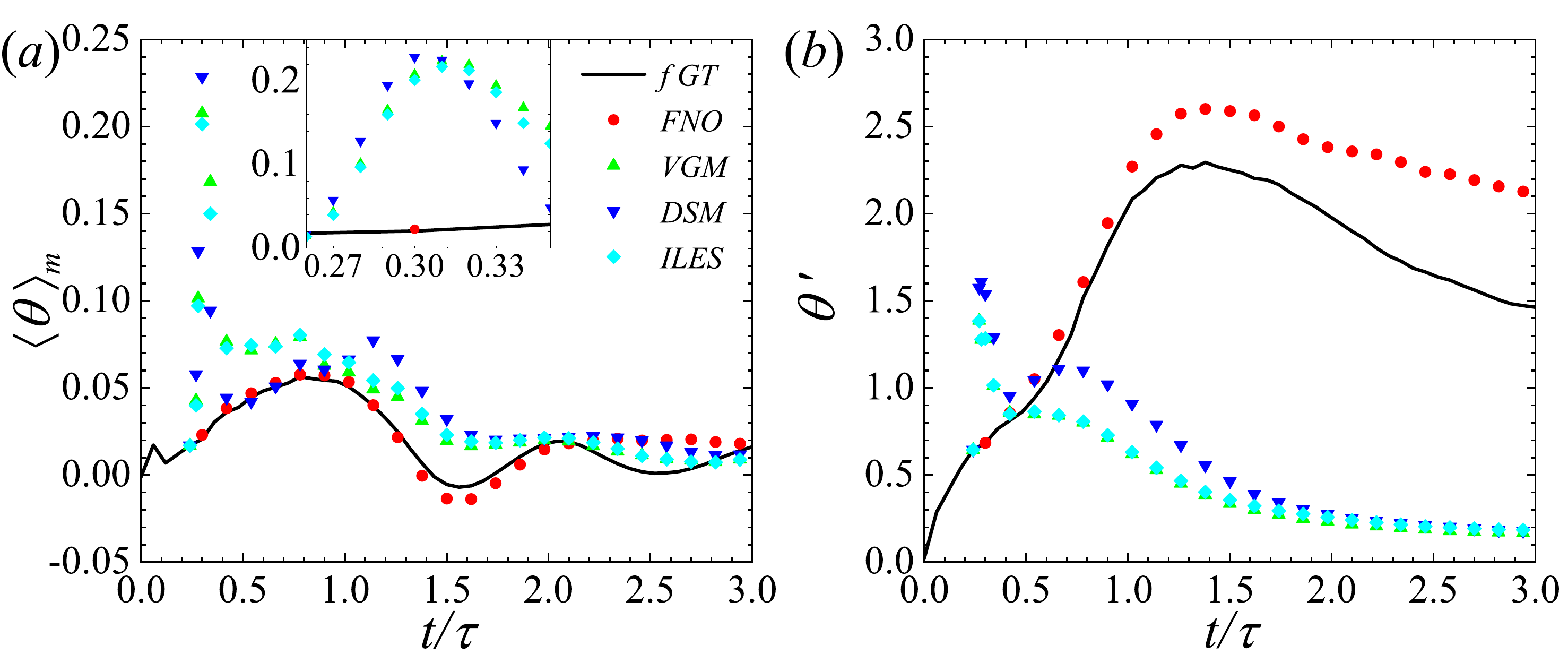} 	 	
		\caption {Time evolution of (a) spatially averaged velocity divergence $\left\langle \theta \right\rangle_m $, inset: the enlarged image at $ 0.25 \leq t/\tau \le 0.35$, and (b) the rms value of velocity divergence $\theta'$ in the mixing region.}\label{fig.theta}
	\end{figure*} 
	Fig. \ref{fig.theta} displays the time evolution of the spatially averaged velocity divergence $\left\langle \theta \right\rangle_m$ and its rms value $\theta'$ in the mixing region. Overall, the FNO model significantly outperforms the ILES, VGM, and DSM models in predicting velocity divergence. The mean velocity divergence $\left\langle \theta \right\rangle_m$ fluctuates towards a positive value, consistent with previous studies\cite{Luo2021,Luo2022}. The FNO model fully captures this trend of fluctuation, while the mean velocity divergence predicted by traditional LESs shows little fluctuation with time. Similar to the $\omega'(0)$, the rms value of velocity divergence $\theta'$ in the mixing layer also peaks around $1.3t/\tau$. The prediction of $\theta'$ by the FNO model coincides with filtered ground truth in the early stages but exhibits some deviation in the later stages. In contrast to the prediction of $\omega'(0)$, the FNO model overestimates $\theta'$. On the other hand, predictions of $\theta'$ by ILES, VGM, and DSM models show significant discrepancies from filtered ground truth, with their rms values nearly ten times lower than filtered ground truth in the later stages. The flow field simulated by the traditional LES method tends to be in a state with zero velocity divergence, which is inconsistent with the ground truth.
	It is worth noting that in the initial stages of LESs, statistical quantities of velocity divergence have large values, which is caused by the subsequent calculation using coarse grids after reading the filtered ground truth flow field at $t/\tau = 0.24$. The enlarged image of $\left\langle \theta \right\rangle_m $ at $ 0.25 \leq t/\tau \le 0.35$ is shown in the inset. The diffusive mixing of the two unequal-density fluids leads to large velocity divergence. \cite{Luo2021,Luo2022,Cook2001} Due to insufficient grid resolution, there is a significant discretization error for velocity divergence at the interface. 
		
	It is worth noting that even though the FNO model was trained using data at $t/\tau \leq 2.0$, satisfactory results were obtained even beyond $t/\tau \geq 2.0$ in the \emph{a posteriori} test, including statistical quantities of velocity and vorticity at $z=0$, as well as velocity divergence in the mixing layer.
	
	\begin{figure*}[ht]\centering
		\setlength{\abovecaptionskip}{0.cm}
		\setlength{\belowcaptionskip}{-0.cm}
		\includegraphics[width=0.8\textwidth]{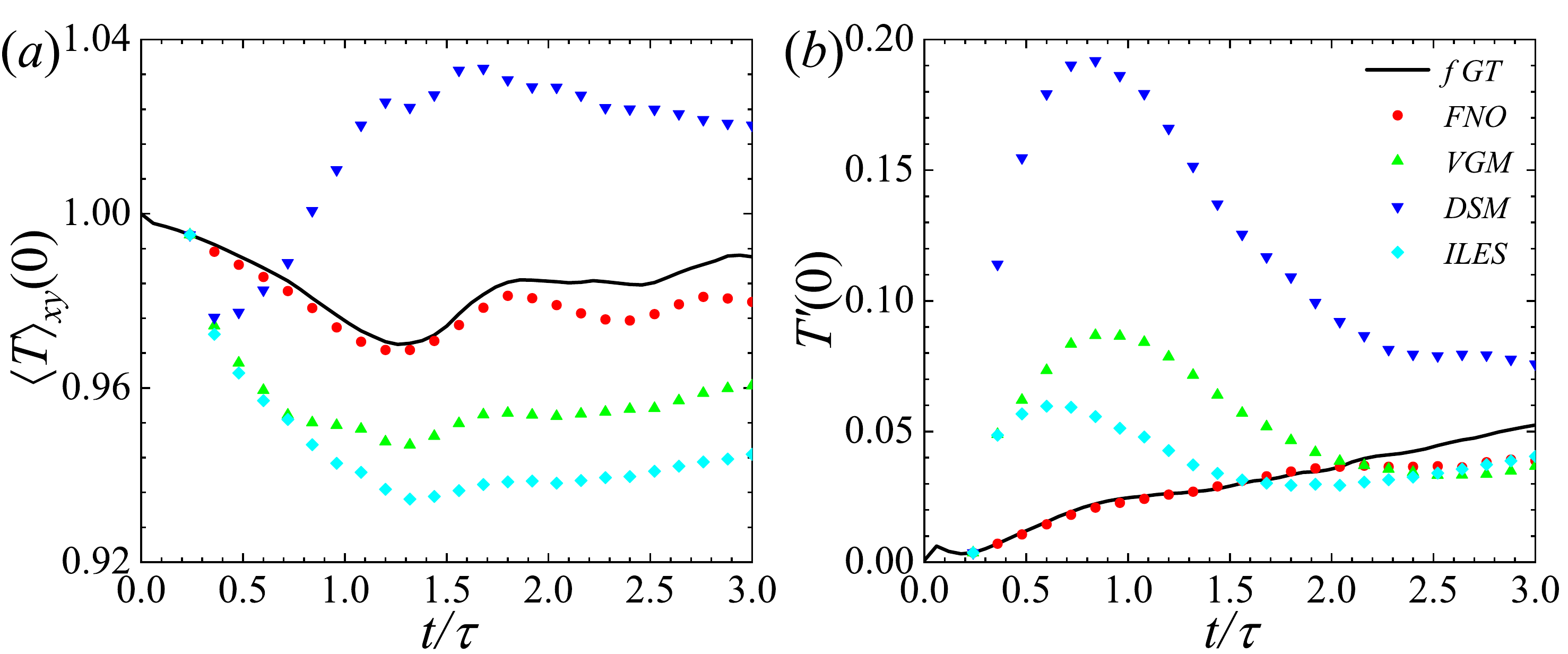} 	 	
		\caption {Time evolution of (a) mean value of temperature, $\left\langle T \right\rangle_{xy} (0) $, and (b) the rms value of temperature, $T'(0)$ at $z=0$.}\label{fig.T}
	\end{figure*}

	\begin{figure*}[ht]\centering
		\setlength{\abovecaptionskip}{0.cm}
		\setlength{\belowcaptionskip}{-0.cm}
		\includegraphics[width=0.8\textwidth]{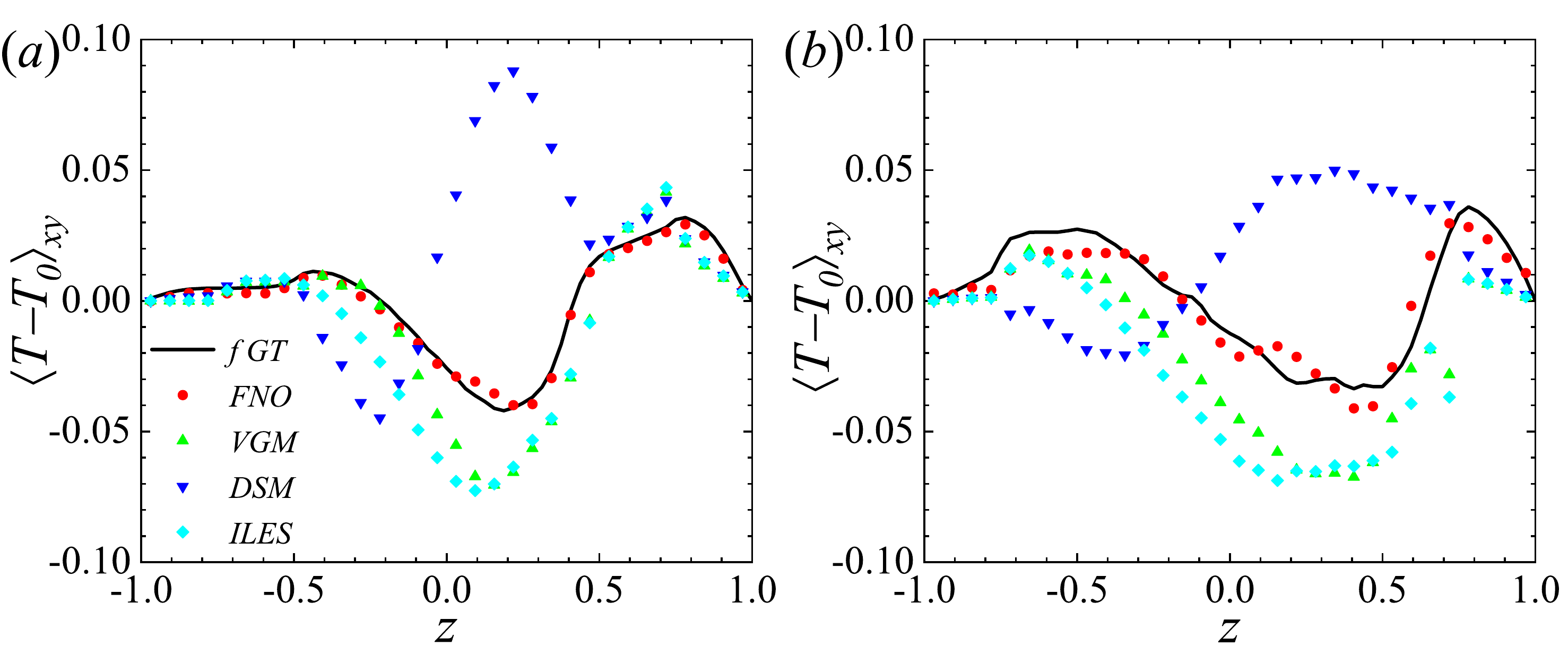} 	 	
		\caption {Mean temperature profiles $\left\langle T-T_0\right\rangle_{xy} $ as a function of $z$ at (a) $t/\tau=1.5$ and (b) $t/\tau=2.7$}\label{fig.pro_T}
	\end{figure*} 

    Fig. \ref{fig.T} presents the time evolution of the mean temperature $\left\langle T \right\rangle_{xy} (0)$ and the rms values of temperature fluctuations $T'(0)$ at $z=0$. Fig. \ref{fig.pro_T} shows the profiles of the mean temperature $\left\langle T-T_0\right\rangle_{xy}$ at $t/\tau=1.5, 2.7$. Due to the weak temperature fluctuations and the difficulty in accurately computing large-scale pressure-dilatation on coarse grids, traditional LESs fail to predict the temperature field accurately. As shown in the figures, although ILES and VGM models predict similar trends in the temporal and spatial variations of mean temperature compared to filtered ground truth, the temperature changes are more drastic than filtered ground truth. Additionally, the DSM model not only predicts drastic temperature changes but also exhibits an opposite trend in temperature variation. However, the FNO model accurately predicts the temperature field, with both the temporal and spatial variations of mean temperature closely matching filtered ground truth.
    
    In contrast to the trend observed in the rms values of other physical quantities, the rms value of temperature fluctuation $T'(0)$ continuously increases over time, except for an initial short period. The prediction ability of FNO model on $T'(0)$ is far superior to that of traditional LESs. ILES and VGM models exhibit significant deviations from filtered ground truth in predicting $T'(0)$ during the early stages, and their predictions gradually approach filtered ground truth as mixing develops. The DSM model performs poorly not only in predicting mean temperature but also in predicting the rms value of temperature fluctuation.

\begin{figure*}[ht]\centering
	\setlength{\abovecaptionskip}{0.cm}
	\setlength{\belowcaptionskip}{-0.cm}
	\includegraphics[width=0.8\textwidth]{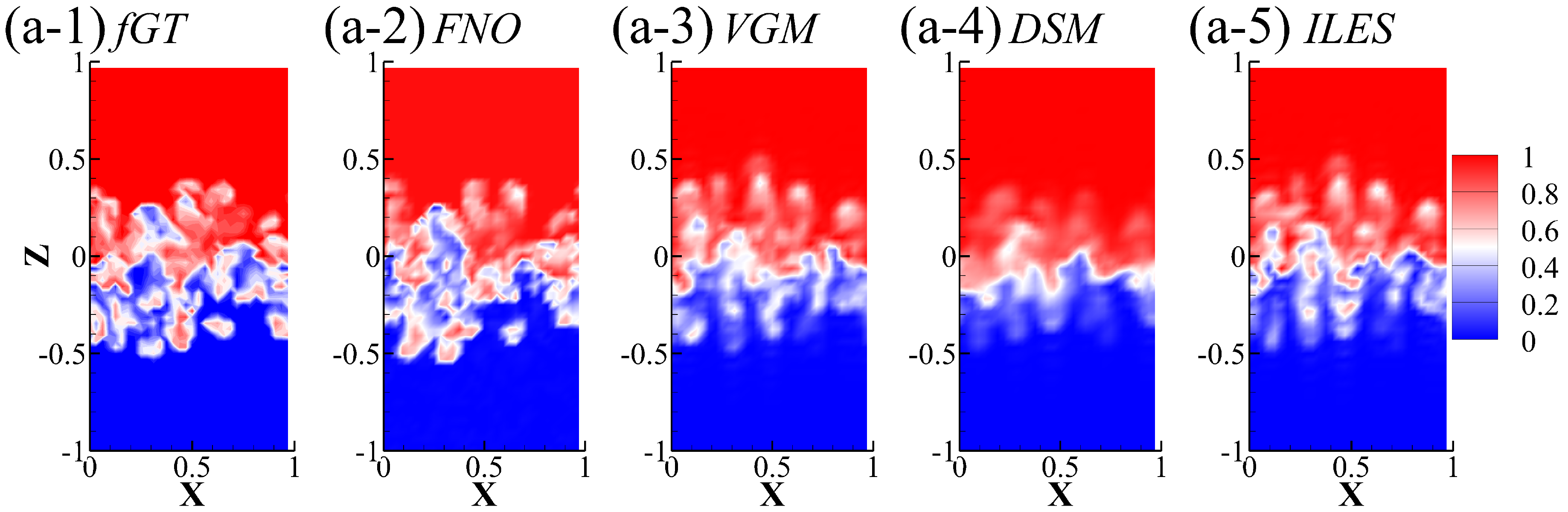}  
	\includegraphics[width=0.8\textwidth]{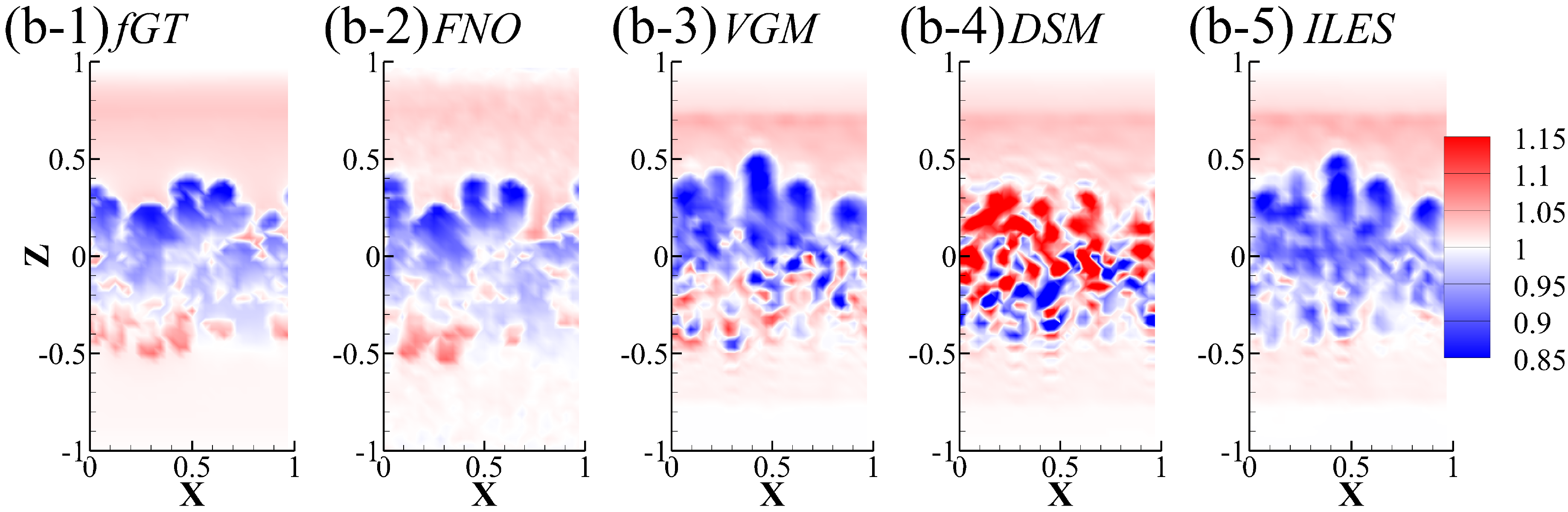}  	 	
	\caption {The contours of (a) concentration $c$ and (b) temperature $T$ at $t/\tau =1.5 $.}\label{fig.con-T}
\end{figure*}  

    
    Furthermore, in Fig. \ref{fig.con-T}, the instantaneous contours of concentration and temperature at $t/\tau=1.5$ are displayed. The temperature and concentration contours predicted by the FNO model are very similar to the contours of filtered ground truth, showing the temperature of the light fluid moving upward decreases, while the temperature of the heavy fluid moving downward increases.\cite{Luo2021,Luo2022} In contrast, other traditional LESs do not exhibit a clear corresponding relationship.
    
    Both the temperature statistics at $z=0$ and the profiles of the mean temperature at $t/\tau=2.7$ indicate that the current FNO model also possesses excellent time generalization capability in predicting the temperature field. We chose the boundary conditions with a fixed temperature, and the profiles of mean temperature align with the temperature boundary conditions at $t/\tau=2.7$. This suggests, to some extent, that the FNO model has learned the form of boundary conditions. For flow fields not used during training ($t/\tau \geq 2.0$), it ensures that the flow field consistently meets the boundary conditions during long-term predictions.  

\subsection{\label{sec:Re} Generalization on higher Reynolds numbers}
	In this section, we will discuss the generalization ability of the FNO model at high Reynolds numbers. The FNO model can efficiently learn large-scale dynamics of turbulence from flow data. Our previous study showed that the FNO has some degree of generalization ability at high Reynolds numbers when predicting 3D HIT\cite{Li2022,Li2023}. Here, we aim to further test the generalization ability of the FNO model in predicting large-scale dynamics of RT turbulence. In RT turbulence, the Reynolds number not only affects the multi-scale turbulent motions in the fully developed stage but also influences the initial diffusion and linear growth stages\cite{Wei2012}. The Reynolds number of flow fields used to train the FNO models is $Re=10000$. When the Reynolds number is sufficiently high and the perturbations are large-scale disturbances, the impact of $Re$ becomes less pronounced in the diffusion and linear growth stages.
	Therefore, the Reynolds number mainly affects the later turbulent stages. The flow fields at $Re=30000$ are simulated using $256^2 \times 512$ grids. In this section, we use the FNO model trained with low Reynolds number $Re=10000$ data to predict and analyze the flow fields at $Re=30000$ using coarse grid $32^2 \times 64$ without additional training or modifications.
	
	\begin{figure*}[ht]\centering
		\setlength{\abovecaptionskip}{0.cm}
		\setlength{\belowcaptionskip}{-0.cm}
		\includegraphics[width=0.8\textwidth]{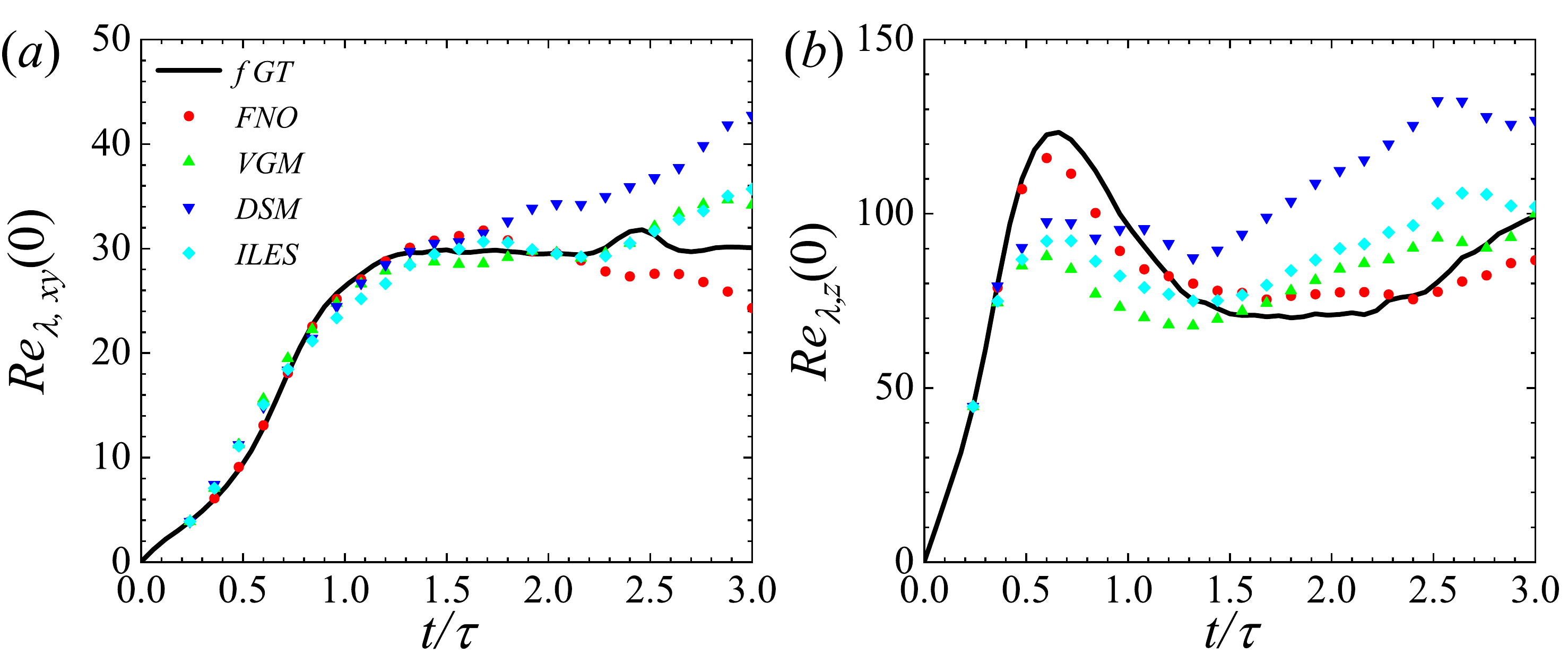} 	 
		\includegraphics[width=0.8\textwidth]{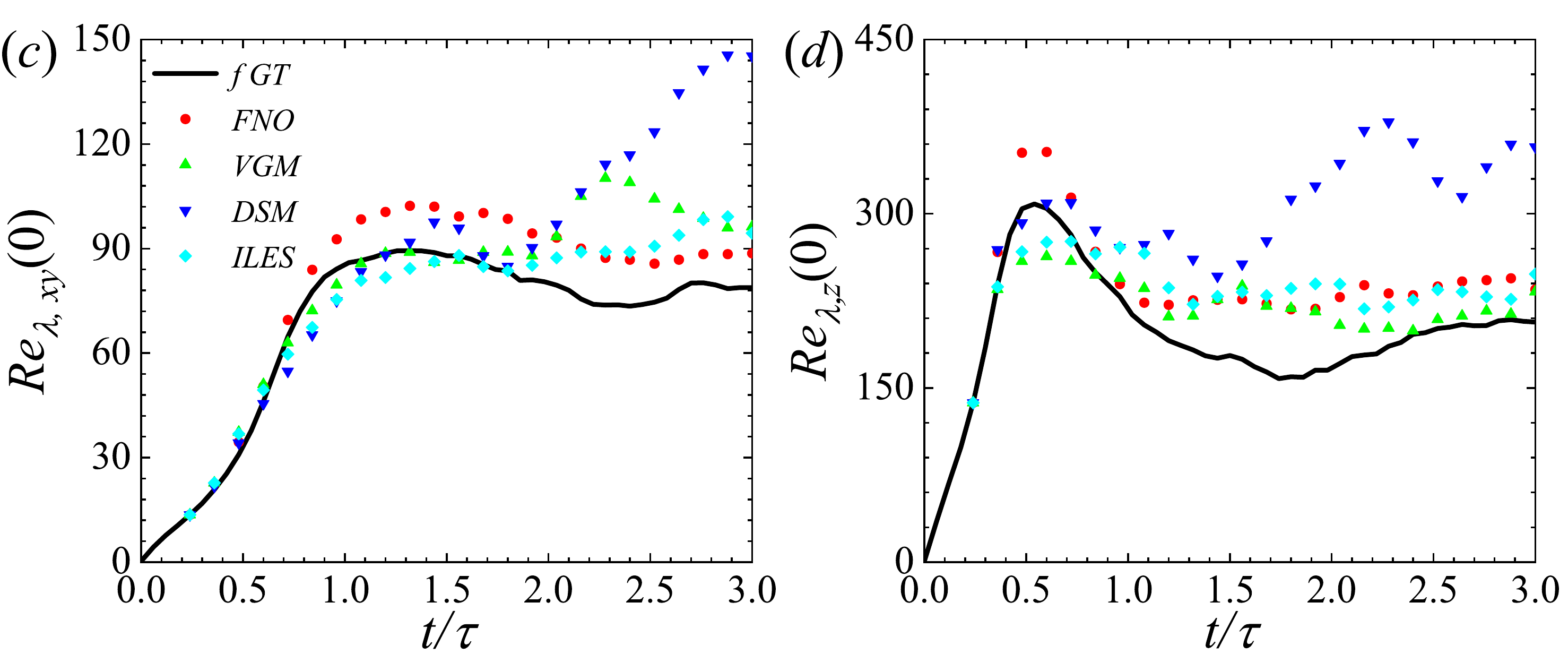} 	 	
		\caption {Time evolution of the horizontal Taylor Reynolds number $R e_{\lambda, xy}(0)$ and vertical Taylor Reynolds number $R e_{\lambda, z}(0)$ at $z=0$: (a) $R e_{\lambda, xy}(0)$ for $Re=10000$, (b) $R e_{\lambda, z}(0)$ for $Re=10000$, (c) $R e_{\lambda, xy}(0)$ for $Re=30000$, (d) $R e_{\lambda, z}(0)$ for $Re=30000$.}\label{fig.Re}
	\end{figure*} 
	The vertical and horizontal Taylor Reynolds numbers at a given vertical coordinate $z$ are respectively defined as \cite{Luo2021,Cook2001}
	\begin{equation}		
		Re_{\lambda, z}(z)=Re \frac{\langle\rho\rangle_{xy} \lambda_{3}\left[\langle 	u_3'^{2}\rangle_{xy}\right]^{1 / 2}}{\mu}   , \quad
		Re_{\lambda, xy}(z)=Re \frac{\langle\rho\rangle_{xy} \lambda_{i}\left[\langle 	u_i'^{2}\rangle_{xy}\right]^{1 / 2}}{\mu} \ (\rm sum \ on \ \textit i =1,2),                   
	\end{equation}
	where $u_i'=u_i-\langle {u}_{i} \rangle_{xy}$ is the fluctuation velocity. The three Taylor microscales are given by $\lambda_{i}(z)= \left[\frac{\langle u_i'^{2}\rangle_{xy}} {\langle\left(\partial u_i' / \partial x_i\right)^2\rangle_{xy}}\right]^{1 / 2}  \ (\rm no \ sum \ on \ \textit{i}, \: and \: \textit{i}=1,2,3)$.
		                
	Fig. \ref{fig.Re} depicts the time evolution of horizontal and vertical Taylor Reynolds numbers at $z=0$ for Reynolds numbers $Re=10000$ and $30000$. The Taylor Reynolds number at $Re=30000$ is more than twice that at $Re=10000$.
	Similar to the rms values of velocities, $R e_{\lambda, z}(0)$ reaches its maximum value earlier than $Re_{\lambda, xy}(0)$. At $Re=10000$, the FNO model accurately predicts the evolution of both horizontal and vertical Taylor Reynolds numbers, outperforming traditional LESs, especially concerning the $R e_{\lambda, z}(0)$. For the case of high Reynolds number $Re=30000$, the FNO model continues to provide reasonably accurate predictions of the Taylor Reynolds numbers. It exhibits the best performance in predicting the trends of Taylor Reynolds numbers over time. The predictions of Taylor Reynolds numbers by the DSM model at both Reynolds numbers are unsatisfactory.
	
	\begin{figure*}[ht]\centering
		\setlength{\abovecaptionskip}{0.cm}
		\setlength{\belowcaptionskip}{-0.cm}
		\includegraphics[width=0.8\textwidth]{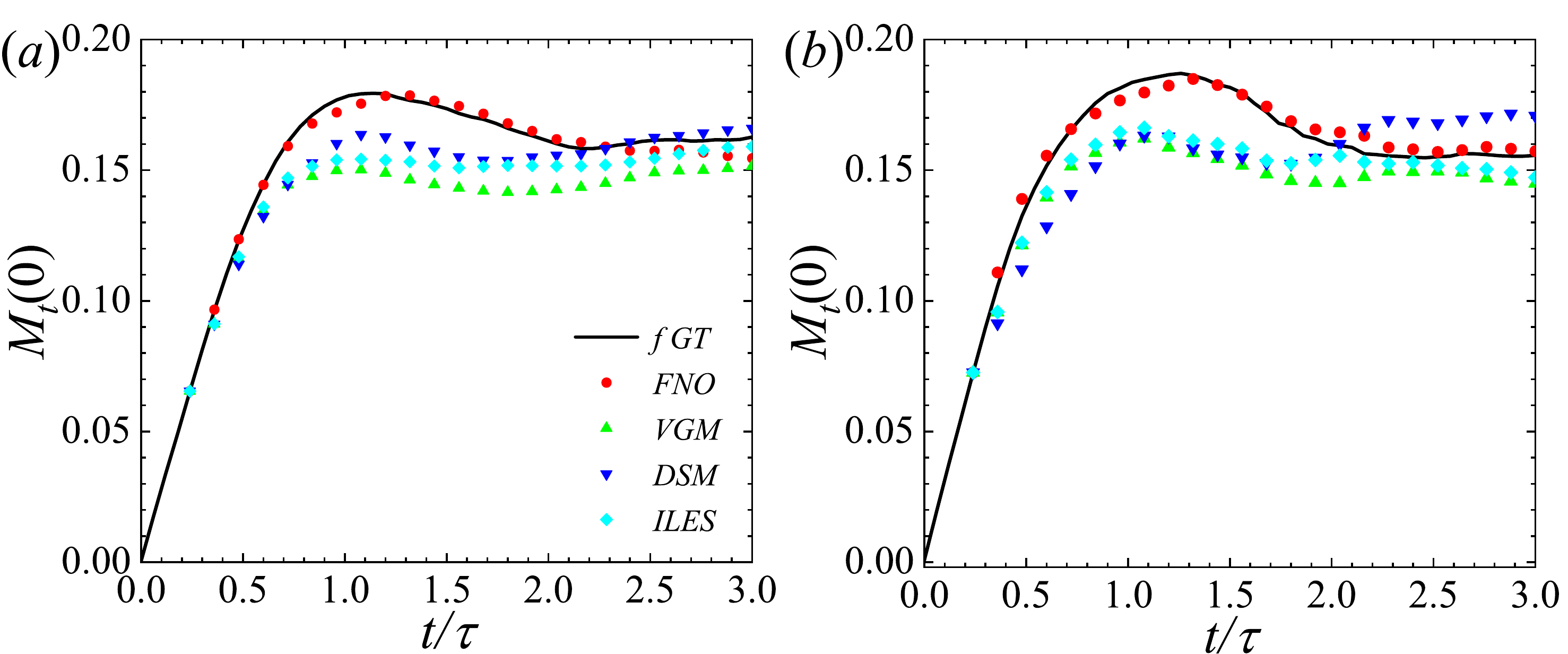} 	 	
		\caption {Time evolution of the turbulent Mach number $M_{t}(0)$ at $z=0$: (a) $Re=10000$ and (b) $Re=30000$.}\label{fig.Mt}
	\end{figure*} 
	The turbulent Mach number is defined as\cite{Luo2021,Luo2022}
	\begin{equation}
		M_{t}(z)=Sr ^{1/2}\frac{\sqrt{\langle u'_i u'_i \rangle_{xy}}}{\langle \gamma p/\rho \rangle_{xy}}.             
	\end{equation}
	Fig. \ref{fig.Mt} illustrates the evolution of turbulent Mach number $M_{t}(0)$ at $z=0$ for $Re=10000, 30000$. At both Reynolds numbers, the turbulent Mach numbers are similar. $M_{t}(0)$ initially experiences rapid growth, followed by a gradual decline, and eventually becomes nearly steady at a constant value. At both Reynolds numbers $Re=10000, 30000$, $M_{t}(0)$ predicted by the FNO model closely matches the filtered ground truth (fGT), even for $t/\tau \geq 2.0$. In contrast, the predictions by the ILES, VGM, and DSM models are worse than those of the FNO model. The increase in Reynolds number mainly affects the vortex motion, without significant influence on the expansion-compression motion. Therefore, the FNO model trained using flow field data at $Re=10000$ exhibits excellent long-term prediction ability at high Reynolds number $Re=30000$.
	
	\begin{figure*}[ht]\centering
		\setlength{\abovecaptionskip}{0.cm}
		\setlength{\belowcaptionskip}{-0.cm}
		\includegraphics[width=0.8\textwidth]{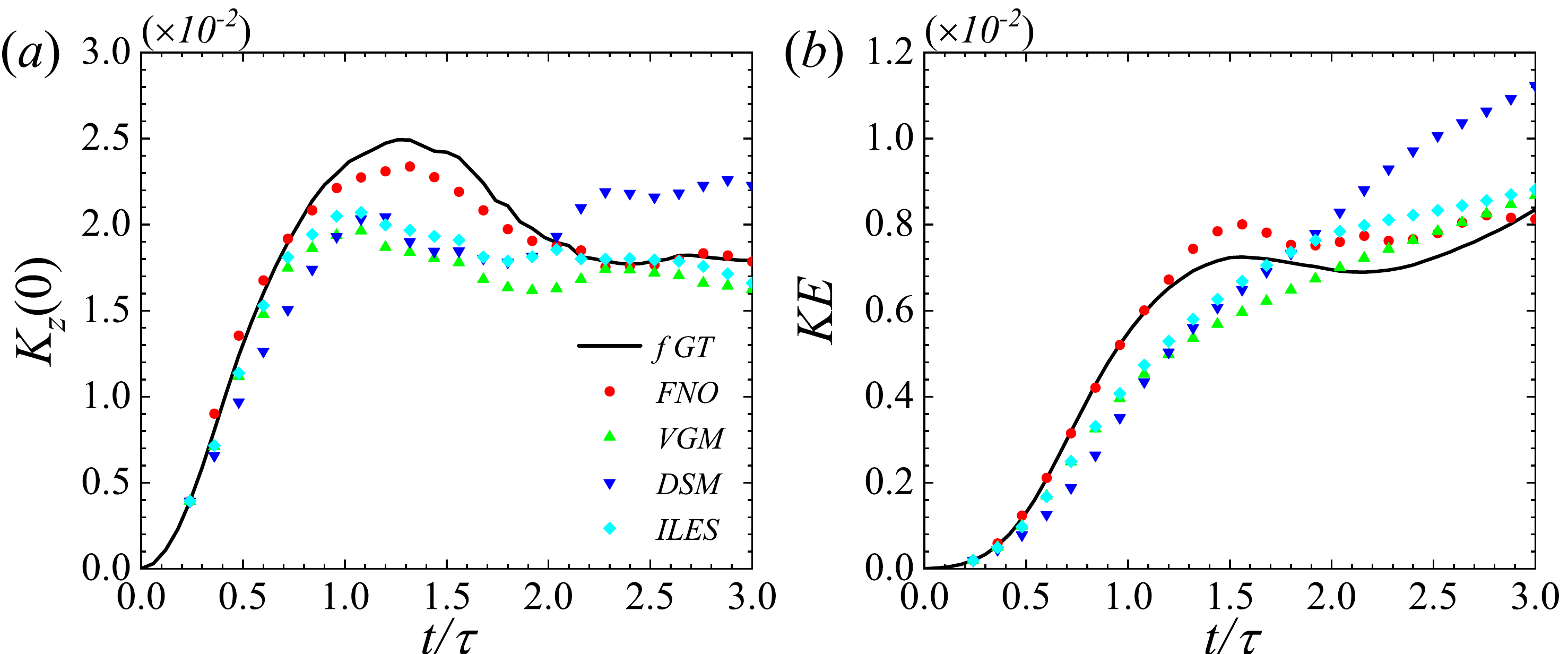} 	 	
		\caption {Time evolution of spatially averaged kinetic energy in the central $x-y$ plane, $K_z(0)$, and in the computational domain, $KE$, for $Re=30000$: (a) $K_z(0)$ and (b) $KE$.}\label{fig.KE-30000}
	\end{figure*} 
	Fig. \ref{fig.KE-30000} shows the evolution of spatially averaged kinetic energy $K_z(0)$ at $z=0$ and the kinetic energy $KE$ in the computational domain for Reynolds number $Re=30000$. It is observed that the FNO model continues to predict the evolution of kinetic energy accurately, outperforming traditional ILES, VGM, and DSM models. Moreover, for the kinetic energy $KE$ in the computational domain, the results of traditional LESs deviate more significantly, especially the DSM model.
	
	\begin{figure*}[ht]\centering
		\setlength{\abovecaptionskip}{0.cm}
		\setlength{\belowcaptionskip}{-0.cm}
		\includegraphics[width=0.8\textwidth]{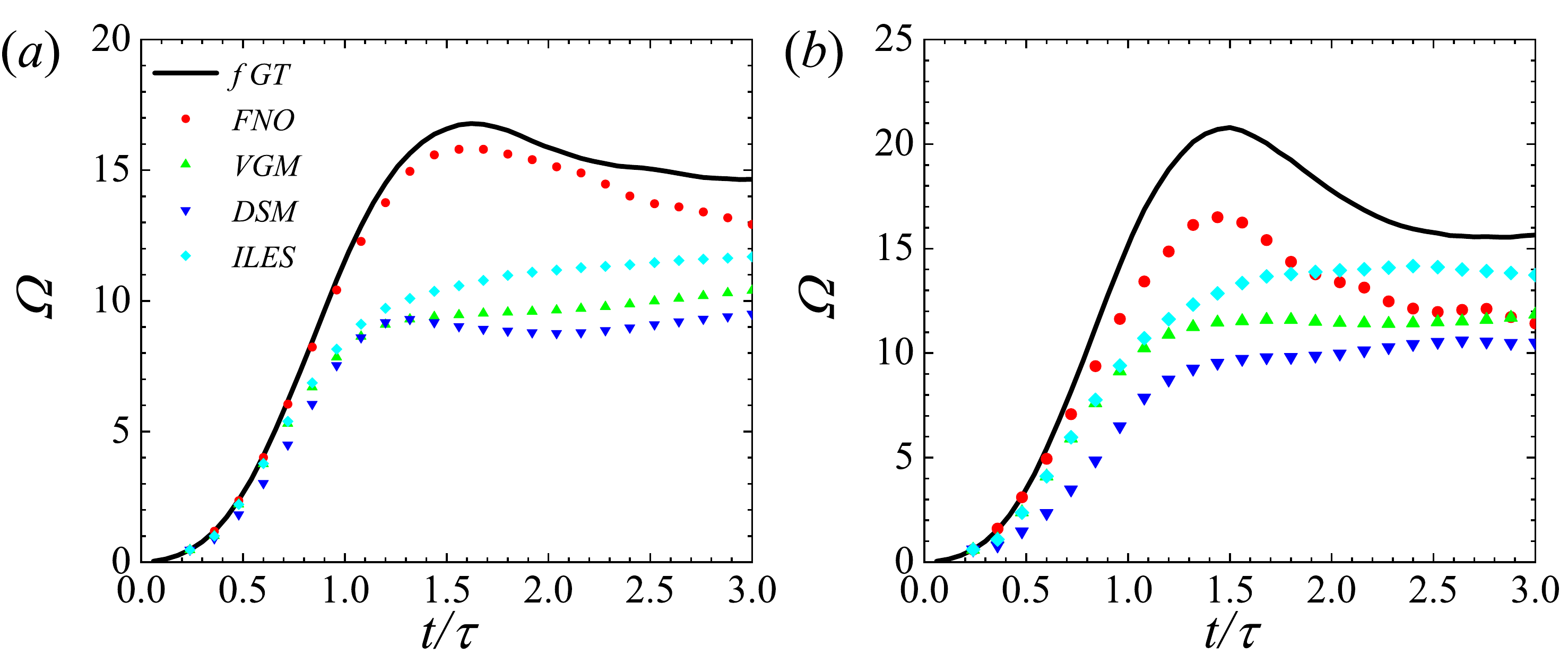} 	 	
		\caption {Time evolution of spatially averaged enstrophy, $\Omega$, in the computational domain: (a) $Re=10000$ and (b)  $Re=30000$.}\label{fig.ENS}
	\end{figure*} 
    Fig. \ref{fig.ENS} depicts the evolution of the enstrophy $\Omega=\left\langle \rho \omega_i^2 /2 \right\rangle$ in the computational domain for $Re=10000, 30000$. At $Re=10000$, the FNO model provides excellent predictive results. However, at high Reynolds number $Re=30000$, although the FNO model captures the trend of $\Omega$ over time, its predictions are lower than the filtered ground truth. Among all models, the FNO model has the best overall agreement with the filtered ground truth at both Reynolds numbers.
    Traditional ILES, VGM, and DSM models only accurately predict the initial growth stage, failing to give a reasonable prediction at about $t/\tau > 1.0 $.
    
    In conclusion, similar to the situation of HIT in our previous study,\cite{Li2022,Li2023} the FNO model also demonstrates certain generalization ability in compressible RT turbulence at high Reynolds numbers. Although its predictive performance is not as good at high $Re$ as at low $Re$, it still outperforms traditional LESs.

\subsection{\label{sec:Times} Computational efficiency of the FNO model}

The FNO model shows its great potential in terms of computational speed in 3D HIT, free shear turbulence, and turbulent channel flow.\cite{Li2022,Li2023,Li2024,Wang2024} We compare the computational efficiency for different models in compressible RT turbulence.
Table \ref{FNO.efficiency} shows the time consumption of 10 FNO steps ($0.6\tau$) using different models. The FNO model is tested on the NVIDIA A100 GPU, where the CPU type is AMD EPYC 7763 @2.45GHz. Moreover, the FNO model is also tested on the CPU, which type is Intel Xeon E5-2690v3 @2.45GHz. The ILES, VGM, and DSM models are simulated using Fortran on the CPU, which is Intel Xeon E5-2690v3 with 8 cores. The time consumption of LESs shown in Table \ref{FNO.efficiency} is the time in numerical simulations without multiplying the number of computation cores. If only one core is used, the simulation time will be longer.
The FNO model using GPU is hundreds or thousands of times faster than traditional LES methods. Even on the CPU, the computational efficiency of the FNO model is much higher than that of the ILES, VGM, and DSM models. 

\begin{table*}
	\caption{\label{FNO.efficiency} Computational cost of 10 FNO steps ($0.6\tau$).}
	\centering
	\setlength\tabcolsep{10pt}
	\begin{tabular}{cccccc}
		\hline\hline
		& FNO & ILES &  VGM & DSM \\
		\hline
		GPU $\cdot$ s & 0.12  & N/A &  N/A & N/A \\
		CPU $\cdot$ s & 17.3 ($\times$ 1 cores)  & 18.8 ($\times$ 8 cores) &  40.2 ($\times$ 8 cores) & 72.0 ($\times$ 8 cores) \\
		\hline\hline
	\end{tabular}
\end{table*}

\section{\label{sec:conclusion}CONCLUSIONS}
In this paper, we apply the Fourier neural operator (FNO) method to the large eddy simulation (LES) of compressible Rayleigh-Taylor (RT) turbulence with miscible fluids. 
Ground truth flow data are generated from numerical simulations of compressible RT turbulence in a $1^2 \times 2$ rectangular box with a grid resolution of $128^2 \times 256$ for Reynolds number $Re=10000$. The training data uses a coarsened flow field with a resolution of $32^2 \times 64$ given by filtering the ground truth flow data. In our numerical studies, the Atwood number is $A_t=0.5$, and the stratification parameter is $Sr=1.0$. The training data is dimensionless physical fields normalized by rms values, which is necessary for compressible flow, due to the different magnitudes of physical fields. 

In the \emph{a posteriori} tests, the results predicted by the FNO model generally converge after approximately 30 training epochs. The data used for training the FNO model only include $t/\tau \leq 2.0$, but we predicted quite good statistical results and instantaneous flow fields until $t/\tau=3.0$ in the \emph{a posteriori} tests, demonstrating generalization ability on longer time prediction of the trained FNO model.

The performances of the FNO model, the traditional VGM and DSM models, and ILES are compared in the \emph{a posteriori} tests of compressible RT turbulence. All statistical results and instantaneous fields predicted by the FNO model are superior to those obtained by traditional ILES, VGM, and DSM models. Particularly, the predictions of velocity divergence and temperature fields by the FNO model are much better than those of other models. Moreover, the computational efficiency of the FNO model is much higher than that of traditional LES methods.

We apply the FNO model trained with Reynolds number $Re=10000$ directly to flow prediction at a higher Reynolds number $Re=30000$, and evaluate the generalization ability of the FNO model at high Reynolds numbers. Overall, although the performance of the FNO model at $Re=30000$ is not as good as that at $Re=10000$, it still outperforms traditional LES methods.

The generalization performance and reliance on DNS data for training are obvious drawbacks of FNO models, which are also present in many machine learning methods. How to improve the generalization performance of the FNO method and reduce its reliance on DNS data is also what we need to do in the future.
In addition, neural network methods have experienced rapid development in recent years, and more advanced models can be attempted for predicting compressible RT turbulence.
	
\section{\label{DATA AVAILABILITY}DATA AVAILABILITY}
The data that support the findings of this study are available from the corresponding author
upon reasonable request.

\begin{acknowledgments}
	This work was supported by the National Natural Science Foundation of China (NSFC
	Grants No. 12172161, No. 92052301, No. 12161141017, and No. 91952104), by the NSFC Basic Science Center Program (Grant No. 11988102), by the Shenzhen Science and Technology Program (Grant No. KQTD20180411143441009), by Key Special Project for Introduced Talents Team of Southern Marine Science and Engineering Guangdong Laboratory (Guangzhou) (Grant No. GML2019ZD0103), and by Department of Science and Technology of Guangdong Province (Grants No. 2019B21203001, No. 2020B1212030001, and No.2023B1212060001). This work was also supported by Center for Computational Science and Engineering of Southern University of Science and Technology, and by National Center for Applied Mathematics Shenzhen (NCAMS).
\end{acknowledgments}

\begin{appendix}
\section{\label{App.512} The comparison between $512^2 \times 1024$, $256^2 \times 512$, and $128^2 \times 256$ grids}
		
\begin{figure*}[ht]\centering
			\setlength{\abovecaptionskip}{0.cm}
			\setlength{\belowcaptionskip}{-0.cm}
			\includegraphics[width=1.0\textwidth]{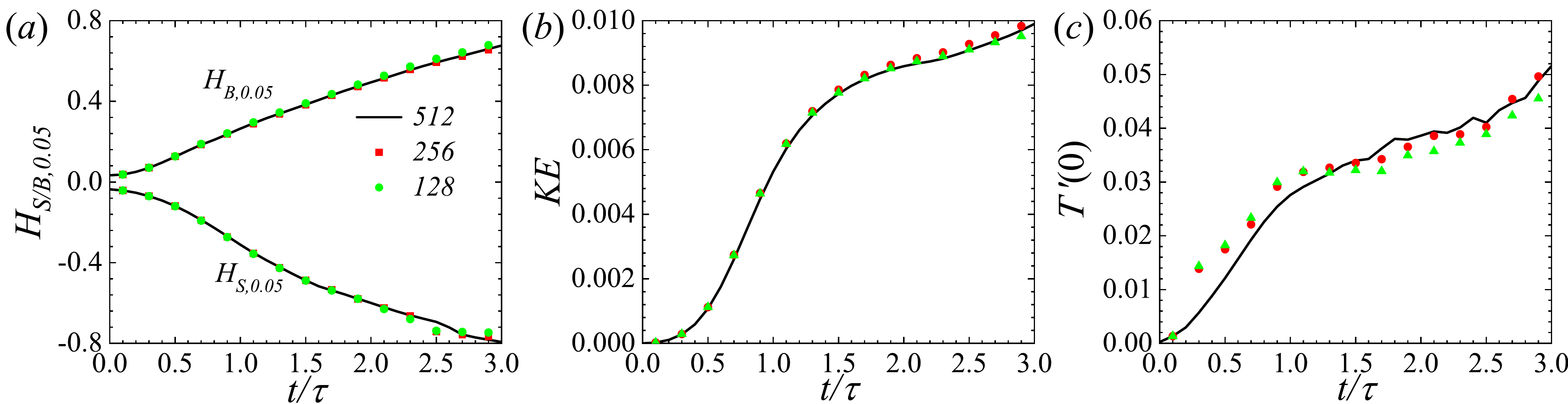} 
			\caption{Statistics in the numerical simulations with $512^2 \times 1024$, $256^2 \times 512$, and $128^2 \times 256$ grids: (a) the evolutions of spike and bubble heights $H_{S/B,0.05}$, (b) the time evolution of spatially averaged kinetic energy in the computational domain $KE$, (c) the time evolution of the rms value of temperature at $z=0$, $T'(0)$.}\label{fig.A1}
\end{figure*}

We compare some statistical results in numerical simulations with $512^2 \times 1024$, $256^2 \times 512$, and $128^2 \times 256$ grids in Fig. \ref{fig.A1}, and differences are not significant. $128^2 \times 256$ grid resolution can reliably predict the compressible RT turbulence and is enough to properly capture the low-order statistics considered in this work.
\end{appendix}

\bibliography{reference}

\end{document}